\documentclass{article}

\usepackage[a4paper]{geometry}

\usepackage[utf8]{inputenc}
\usepackage{enumerate}

\usepackage[english]{babel}

\usepackage{graphicx}
\graphicspath{{.}{}{figs/}{figures/}{images/}} 
\DeclareGraphicsExtensions{.pdf,.png,.jpg}

\usepackage{subfigure}
\usepackage{url}

\usepackage{tabularx} 
\usepackage{multirow} 
\usepackage{rotating} 

\usepackage{amsmath}
\usepackage{amsfonts}
\usepackage{amssymb}  
\usepackage{amstext}
\usepackage{wasysym}
\usepackage{upref}
\usepackage{MnSymbol} 
\usepackage[amsmath,thmmarks,thref]{ntheorem}

\newtheorem{theorem}{Theorem}[section] 

\newtheorem{lemma}[theorem]{Lemma}

\theoremstyle{plain}
\newtheorem{definition}[theorem]{Definition}

\theoremstyle{plain}

\theoremstyle{nonumberplain}
\theoremheaderfont{\normalfont\itshape}
\theorembodyfont{\normalfont}
\theoremseparator{.}
\theoremsymbol{$\square$}

\numberwithin{equation}{section}

\usepackage{microtype}

\title{A primer on information theory, with applications to neuroscience}
\author{Felix Effenberger\\[2ex]Max-Planck-Institute for Mathematics in the Sciences, Leipzig, Germany}
\date{April 2013}

\begin{document}

\enlargethispage{1cm}
\maketitle

\tableofcontents

\thispagestyle{empty}

\clearpage
\begin{abstract}
Given the constant rise in quantity and quality of data obtained from
neural systems on many scales ranging from molecular to systems',
information-theoretic analyses became increasingly necessary during
the past few decades in the neurosciences. Such analyses can provide
deep insights into the functionality of such systems, as well as a
rigid mathematical theory and quantitative measures of information
processing in both healthy and diseased states of neural systems. This
chapter will present a short introduction to the fundamentals of
information theory, especially suited for people having a less firm
background in mathematics and probability theory. To begin, the
fundamentals of probability theory such as the notion of probability,
probability distributions, and random variables will be reviewed.
Then, the concepts of information and entropy (in the sense of
Shannon), mutual information, and transfer entropy (sometimes also
referred to as conditional mutual information) will be outlined. As
these quantities cannot be computed exactly from measured data in
practice, estimation techniques for information-theoretic quantities
will be presented. The chapter will conclude with the applications of
information theory in the field of neuroscience, including questions
of possible medical applications and a short review of software
packages that can be used for information-theoretic analyses of neural
data.
\end{abstract}

\clearpage
\section{Introduction}
\label{sec:introduction}

Neural systems process information. This processing is of fundamental biological importance for all animals and humans alike as its main (if not sole) biological purpose is to ensure the survival of an individual (in the short run) and its species (in the long run) in a given environment by means of perception, cognition, action and adaption.

Information enters a neural system in form of sensory input representing some aspect of the outside world, perceivable by the sensory modalities present in the system. After processing this information or parts of it, the system may then adjust its state and act according to a perceived  change in the environment. 

This general model is applicable to very basic acts of cognition as well as to ones requiring higher degrees of cognitive processing. Yet, the underlying principle is the same. Thus measuring, modeling and (in the long run) understanding information processing in neural systems is of prime importance for the goal of gaining insight to the functioning of neural systems on a theoretical level.   

Note that this question is of theoretical and abstract nature so that we take an abstract view on information in what follows. We use Shannon's theory of information \cite{Shannon1948} as a tool that provides us with a rigid mathematical theory and quantitative measures of information. Using information theory, we will have a conceptual look at information in neural systems. In this context, information theory can provide both explorative and normative views on the processing of information in a neural system as we will see in  Section~\ref{sec:neuralsystems}. In some cases, it is even possible to gain insights on the nature of the ``neural code'', i.e.\ the way neurons transmit information via their spiking activity.

Information theory was originally used to analyze and optimize man-made communication systems, for which the functioning principles are known. None the less, it was soon realized that the theory could also be used in a broader setting, namely to gain insight into the functioning of systems for which the underlying principles  are far from fully understood, such as neural systems for example. This was the beginning of the success story of information-theoretic methods in many fields of science such as economics, psychology, biology, chemistry, physics and many more.

The idea of using information theory to quantitatively assess information processing in neural systems has been around since the 1950s, see the works of Attneave \cite{Attneave1954}, Barlow \cite{Barlow1961} and Eckhorn and Pöpel \cite{Eckhorn1974, Eckhorn1975}. Yet, as information-theoretic analyses are data-intensive, these methods were rather heavily restricted by (a) the limited resources of computer memory and computational power available and (b) the limited accuracy and amount of measured data that could be obtained from neural systems (on the single cell as well as at the systems level) at that time. However, given the constant rise in available computing power and the evolution and invention of data acquisition techniques that can be used to obtain data from neural systems (such Magnetoencephalography (MEG), functional magnetic resonance imaging (fMRI) or calcium imaging), information-theoretic analyses of all kinds of biological and neural systems became more and more feasible and could be carried out with greater accuracy and for larger and larger (sub-)systems. 

Over the last decades such analyses became possible using an average workstation computer, a situation that could only be dreamed of in the 1970s. Additionally, the emergence of new non-invasive data-collection methods such as fMRI and MEG that outperform more traditional methods like Electroencephalography (EEG) in terms of spatial resultion (fMRI, MEG) or noise-levels (MEG) made it possible to even obtain and analyze system-scale data of the human brain in vivo. 


The goal of this chapter is to give a short introduction to the fundamentals of information theory and its application to data analysis problems in the neurosciences. And although information-theoretic analyses of neural systems were not often used in order to gain insight on or characterize neural dysfunction so far, this could prove to be a helpful tool in the future.

The chapter is organized as follows. We first talk a bit about the process of modeling in Section~\ref{sec:setting} that is fundamental for all what follows as it connects reality with theory. As information theory is fundamentally based on probability theory, following this we give an introduction to the mathematical notions of probabilities, probability distributions and random variables in Section~\ref{sec:probabilities}. If you are familiar with probability theory, you may well skim or skip this section. Section~\ref{sec:informationentropy} deals with the main ideas of information theory. We first take a view on what we mean by information and introduce the core concept of information theory, namely \emph{entropy}. Starting from the concept of entropy, we will then continue to look at more complex notions such as \emph{conditional entropy} and \emph{mutual information} in Section~\ref{sec:mutualinformation}. We will then consider a variant of \emph{conditional mutual information} called \emph{transfer entropy} in Section~\ref{sec:transferentropy}. We conclude the theoretical part by discussing methods used for the estimation of information-theoretic quantities from sampled data in Section~\ref{sec:estimation}. What follows will deal with the application of the theoretical measures to neural data. We then give a short overview of applications of the discussed theoretical methods in the neurosciences in Section~\ref{sec:neuralsystems}, and last (but not least), Section~\ref{sec:software} constrains a list of software packages that can be used to estimate information theoretic quantities for some given data set.

\section{Modeling}
\label{sec:setting}

In order to analyze the dynamics and gain a theoretical understanding of a given complex system, one usually defines a model first, i.e.\ a simplified theoretical version of the system to be investigated. The rest of the analysis is then based on this model and can only capture aspects of the system that are also contained in the model. Thus, care has to be taken when creating the model as the following analysis crucially depends on the quality of the model.


When building a model based on measured data, there is an important thing we have to pay attention to, namely that any data obtained by measurement of physical quantities is only accurate up to a certain degree and corrupted by noise. This naturally also holds for neural data (e.g.\ electrophysiological single- or multi-cell measurements, EEG, fMRI or MEG data). Therefore, when observing the state of some system by measuring it, one can only deduce the true state of the system up to a certain error determined by the noise in the measurement (which may depend both on the measurement method and the system itself). In order to model this uncertainty in a mathematical way, one uses probabilistic models for the states of the measured quantities of a system.  This makes probability theory a key ingredient to many mathematical models in the natural sciences.

\section{Probabilities and Random Variables}
\label{sec:probabilities}

The roots of the mathematical theory of probability lie in the works of Cardano, Fermat, Pascal, Bernoulli and de Moivre in the 16th and 17th century, in which the authors attempted to analyze games of chance. Pascal and Bernoulli were the first to treat the subject as a branch of mathematics, see \cite{Todhunter05HistMathTheoProb} for a historical overview. Mathematically speaking, probability theory is concerned with the analysis of random phenomena. Over the last centuries, it has become a well-established mathematical subject. For a more in-depth treatment of the subject see \cite{Hoel71IntProbTheo, Shiryayev84Prob, Klenke08ProbTheo}.


\subsection{A First Approach to Probabilities via Relative Frequencies}

Let us consider an experiment that can produce a certain fixed number of outcomes (say a coin toss, where the possible outcomes are heads or tails or the throw of a die where the die will show one of the numbers $1$ to $6$). The set of all possible outcomes is called the \emph{sample space} of the experiment.

One possible result of an experiment is called \emph{outcome} and a set of outcomes is called an \emph{event} (for the mathematically adept: an event is a subset of the power set of all outcomes). Take for example the throw of a regular, $6$-sided die as an experiment. The set of results in this case would be the set of natural numbers $\{1,\dots,6\}$ and examples of events are $\{1,3,5\}$ or $\{2,4,6\}$ corresponding to the events ``an odd number was thrown'' and ``an even number was thrown'', respectively.

The classical definition of the probability of an event is due to Laplace: ``The probability of an event to occur is the number of cases favorable for the event divided by the number of total outcomes possible'' \cite{Todhunter05HistMathTheoProb}.

We thus assign each possible outcome a \emph{probability}, a real number between $0$ and $1$ that is thought of as to describe how ``likely'' it is that the given event will occur, where $0$ means ``the event doesn't ever occur'' and $1$ means ``the event always occurs''. The sum of all the assigned numbers is restricted to be $1$ as we assume that one of our considered events always occurs. For the coin toss, the possible outcomes heads and tails thus each have probability $\frac{1}{2}$ (considering that the number of favourable outcomes is $1$ and the number of possible outcomes is $2$) and for the throw of a die this number is $\frac{1}{6}$ for each digit. This assumes that we have a so-called \emph{fair} coin or die, i.e.\ one that does not favor any particular outcomes over the others.  

The probability of a given event to occur is then just the sum of the probabilities of the outcomes the event is composed of, e.g.\ when considering the throw of a die, the probability of the event ``an odd number is thrown'' is $\frac{1}{6}+\frac{1}{6}+\frac{1}{6}=\frac{1}{2}$.

Such types of experiments in which all possible outcomes have the same probability (they are called equiprobable) are called \emph{Laplacian experiments}. The simplest case of an experiment not having equiprobable outcomes is the so called \emph{Bernoulli experiment}. Here, two possible outcomes ``success'' and ``failure'', with probabilities $p\in [0,1]$ and $1-p$ are considered. Let us now consider probabilities in the general setting. 

\subsection{An Axiomatic Description of Probabilities}

The foundations of modern probability theory were laid by Kolmogorov~\cite{Kolmogoroff73GrundbWahrsch} in the 1930s. He was the first to give an axiomatic description of probability theory based on measure theory, putting the field on a mathematically sound basis.  We will state his axiomatic description of probabilities in the following. This rather technical approach might seem a little complicated and cumbersome first and we will try to give well-understandable explanations of the concepts and notions used as they are of general importance.

Kolmogorov's definition is based on what is known as measure theory, a field of mathematics that is concerned with measuring the (geometric) size of subsets of a given space. Measure theory gives an axiomatic description of a \emph{measure} (as a function $\mu$ assigning a non-negative number to each subset) that fulfills the usual properties of a geometric measure of length (in $1$-dimensional space), area (in $2$-dimensional space), volume (in $3$-dimensional space), and so on. For example, if we take the measure of two disjoint (i.e.\ non-overlapping) sets, we expect the measure of their union to be the sum of the measures of the two sets and so on.        

One prior remark on the definition: When looking at sample spaces (remember, these are the sets of possible outcomes of a random experiment) we have to make a fundamental distinction between \emph{discrete sample spaces} (i.e.\ ones in which the outcomes can be separated and counted, like in a pile of sand, where we think of each little sand particle representing one possible outcome) and \emph{continuous sample spaces} (where the outcomes form a continuum and cannot be separated and counted, think of this sample space as some kind of dough in which the outcomes cannot be separated). Although in most cases the continuous setting can be treated as a  straightforward generalization of the discrete case and we just have to replace sums by integrals in the formulas, some technical subtleties exist, that makes a  distinction between the two cases necessary. This is why we separate the two cases in all of what follows.  

\begin{definition}[measure space and probability space]
	\label{def:probspace}
	A \emph{measure space} is a triple $(\Omega,\mathcal{F},\mu)$. Here
	
	\begin{itemize}
	 		\item the \emph{base space} $\Omega$ denotes an arbitrary nonempty set,
	 		\item $\mathcal{F}$ denotes the set of \emph{measurable sets} in $\Omega$ which has to be a so called \emph{$\sigma$-algebra} over $\Omega$, i.e.\ it has to fulfill
	 		\begin{enumerate}[(i)]
	 			\item $\emptyset\in \mathcal{F}$
	 			\item $\mathcal{F}$ is closed under complements: if $E\in\mathcal{F}$, then $(\Omega \backslash E)\in \mathcal{F}$,
	 			\item $\mathcal{F}$ is closed under countable unions: if $E_i\in \mathcal{F}$ for $i=1,2,\dots$, then $(\cup_{i} E_i)\in \mathcal{F}$,
	 		\end{enumerate}
	 		\item $\mu$ is the so called \emph{measure}: It is a function $\mu: \mathcal{F}\rightarrow \mathbb{R}\cup\{\infty \}$ with the following properties
	 		\begin{enumerate}[(i)]
	 			\item $\mu(\emptyset)=0$ and $\mu\geq 0$ (non-negativity),
	 			\item $\mu$ is countably additive: if $E_i\in \mathcal{F}$, $i=1,2,\dots$ is a collection of pairwise disjoint (i.e.\ non-overlapping) sets, then $\mu(\cup_{i} E_i)=\sum_{i} \mu(E_i)$.
	 		\end{enumerate}
	\end{itemize}

\end{definition}

Why this complicated definition of measurable sets, measures, etc.? Well, this is mathematically the probably (no pun intended) most simple way to formalize the notion of a ``measure'' (in terms of geometric volume) as we know it over the real numbers. 

When defining a measure, we first have to fix the whole space in which we want to measure. This is the base space $\Omega$. $\Omega$ can be any arbitrary set: The sample space of a random experiment, e.g.\ $\Omega=\{$heads,tails$\}$ when we look at a coin toss or $\Omega=\{1,\dots,6\}$ when we look at the throw of a die (these are two examples of discrete sets), the set of real numbers $\mathbb{R}$, the real plane $\mathbb{R}^2$ (these are two examples of continuous sets) or whatever you choose it to be. When modeling the spiking activity of a neuron the two states could be ``neuron spiked'' or ``neuron didn't spike''.

In a second step we choose a collection of subsets of $\Omega$ that we name $\mathcal{F}$, the collection of subsets of $\Omega$ that we want to be measurable. Note that the measurable subsets of $\Omega$ are not given a priori, but that we determining those by choosing $\mathcal{F}$. 
So, you may ask, why this complicated setup with $\mathcal{F}$, why not make every possible subset of $\Omega$ measurable, i.e.\ make $\mathcal{F}$ the power set of $\Omega$ (the power set is the set of all subsets of $\Omega$)? This is totally reasonable and can easily been done when the number of elements of $\Omega$ is finite. But as with many things in mathematics, things get complicated when we deal with the continuum: In many natural settings, e.g.\ when $\Omega$ is a continuous set, this is just not possible or desirable for technical reasons. That is why we choose only a subset of the power set (you might refer to its elements as the ``privileged'' subsets) and make only the contained subsets measurable. We want to choose this subset in a way that the usual constructions that we know from geometric measures still work in the usual way, though. This motivates the properties that we impose on $\mathcal{F}$: We expect to be able to measure the complements of measurable sets, as well as the union and intersection of a finite number of measurable sets to again be measurable. These properties are motivated by the corresponding properties of geometric measures (i.e.\ the union, intersection and complement of intervals of certain lengths has a length and so on). So to sum up, the set $\mathcal{F}$ is a subset of the power set of $\Omega$, and sets that are not in $\mathcal{F}$ are not measurable.

In a last step, we choose a function $\mu$ that assigns a measure (think of it as a generalized geometric volume) to each measurable set (i.e.\ each element of $\mathcal{F}$), where the measure has to fulfill some basic properties that we know from geometric measures: The measure is non-negative, the empty set (that is contained in every set) should have measure $0$ and the measure is additive.

All together, this makes the triple $(\Omega,\mathcal{F},\mu)$ a space in which we can measure events and use constructions that we know from basic geometry. Our definition makes sure that the measure $\mu$ behaves in the way we expect it to (mathematicians call this a natural construction). Take some time to think about it: Definition~\ref{def:probspace} above generalizes the notion of the geometric measure in terms of the length $l(I)=b-a$ of intervals $I=[a,b]$ over the real numbers.

In fact, when choosing the set $\Omega=\mathbb{R}$ we can construct the so called \emph{Borel $\sigma$-algebra} $\mathcal{B}$ that contains all closed intervals $I=[a,b]$, $a<b$ and a measure $\mu_{\mathcal{B}}$ that assigns each interval $I=[a,b]\in \mathcal{B}$ its length $\mu_{\mathcal{B}}(I)=b-a$. The measure $\mu_{\mathcal{B}}$ is called \emph{Borel measure}. It is the standard measure of length that we know from geometry and makes $(\mathbb{R},\mathcal{B},\mu_{\mathcal{B}})$ a measure space. This construction can easily be extended to arbitrary dimensions (using closed sets) resulting in the measure space $(\mathbb{R}^n,\mathcal{B}^n,\mu_{\mathcal{B}^n})$ that fulfills the properties of a $n$-dimensional geometric measure of volume.  

Let us look at some examples of measure spaces now: 

\begin{enumerate}
	\item Let $\Omega = \{0,1\}$, $\mathcal{F} = \{\emptyset, \{0\}, \{1\}, \Omega \}$ and $P$ with $P(0)=P(1)=0.5$. This makes $(\Omega, \mathcal{F}, P)$ a measure space for our coin toss experiment. Note that in this simple case, $\mathcal{F}$ equals the full power set of $\Omega$. 
	\item Let $\Omega = \{a, b, c, d\}$ and let $\mathcal{F} = \{\emptyset, \{a, b\}, \{c, d\}, \Omega \}$ with $P(\{a, b\})=p$ and $P(\{c, d\})=1-p$, where $p$ denotes an arbitrary number between $0$ and $1$. This makes $(\Omega, \mathcal{F}, P)$ a measure space.
\end{enumerate}

Having understood the general case of a measure space, defining a probability space and a probability distribution is easy. 

\begin{definition}[probability space, probability distribution]
	A \emph{probability space} is a measure space $(\Omega,\mathcal{F},\mu)$ for which the measure $\mu$ is normed, i.e.\ $\mu: \Omega\rightarrow [0,1]$ with $\mu(\Omega)=1$. The measure $\mu$ is called \emph{probability distribution} and is often also denoted by $P$ (for probability). $\Omega$ is called the \emph{sample space}, elements of $\Omega$ are called \emph{outcomes} and $\mathcal{F}$ is the set of \emph{events}.
\end{definition}

Note that again, we make the distinction between discrete and continuous sample spaces here. In the course of history, a probability distribution on a discrete sample space came to be called \emph{probability mass function} (or \emph{pmf}) and a probability distribution defined on a continuous sample space came to be called \emph{probability density function} (or \emph{pdf}).


Let us look at a few examples, where the probability spaces in the following are given by the triple $(\Omega, \mathcal{F}, P)$.

\begin{enumerate}
	\item Let $\Omega = \{$heads,tails$\}$ and let $\mathcal{F} = \{\emptyset,	\{$heads$\}, \{$tails$\}, \Omega \}$. This is a probability  space for our coin toss experiment, where $\emptyset$ relates to the event ``neither heads nor tails'' and $\Omega$ to the event ``either heads or tails''. Note that in this simple case, $\mathcal{F}$ equals the full power set of $\Omega$. 
	
	
	
	\item Let $\Omega = \{1, \dots, 6\}$ and let $\mathcal{F}$ be the full power set of $\Omega$ (i.e.\ the set of all subsets of $\Omega$, there are $6^2=36$, can you enumerate them all?). This is a probability for our experiment of dice throws, where we can distinguish all possible events.
\end{enumerate}

\subsection{Theory and Reality}

\begin{figure}
\centering
	\includegraphics{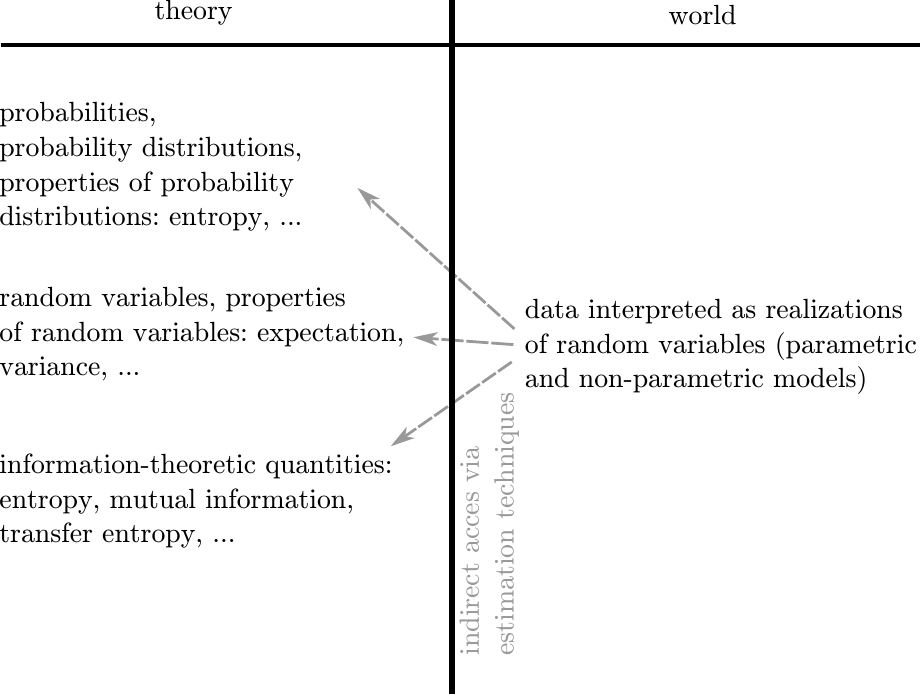}
\caption{Theoretical quantities and measurable quantities: The only things observable and accessible usually are data (measured or generated), all theoretical quantities are not directly accessible. They have to be estimated using statistical methods.}
\label{fig:theoryworld}
\end{figure}

It is important to stress that probabilities themselves are a mathematical and  purely theoretical construct to help in understanding and analyzing random experiments, and per se they do not have to do anything with reality. They can be understood as an ``underlying law'' that generates the outcomes of a random experiment and \emph{can never} be directly observed, see Figure~\ref{fig:theoryworld}. But with some restrictions they can be estimated for a certain given experiment by looking at the outcomes of many repetitions of that experiment.

\begin{figure}
\centering
\subfigure[]{%
\label{fig:diefreq1}%
\includegraphics{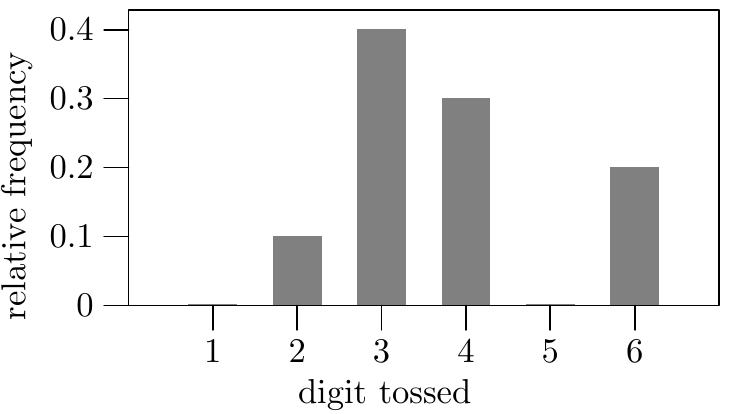}
}
\subfigure[]{%
\label{fig:diefreq2}%
\includegraphics{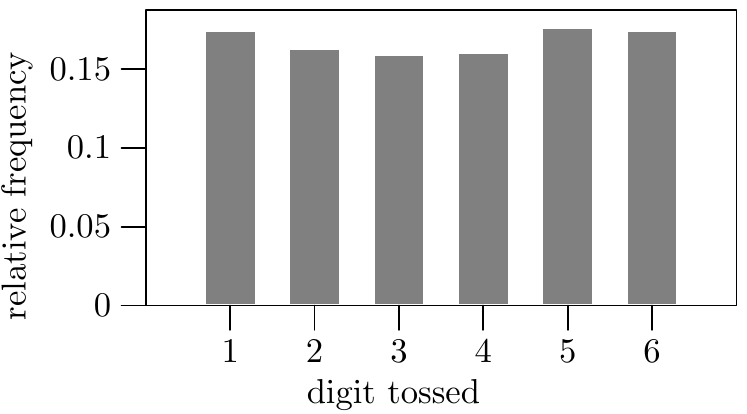}
}
\caption{Relative frequencies of tossed digits using a fair die: \subref{fig:diefreq1} after $10$ tosses and \subref{fig:diefreq2} after $1000$ tosses.}
\label{fig:diefreq}
\end{figure}

Let us consider the following example. Assume that our experiment is the roll of a six-sided die. When repeating the experiment for $10$ times (also called \emph{trials}) we will obtain frequencies for each of the numbers as given in Figure~\ref{fig:diefreq1}. Repeating the experiment for $100$ times we will get frequencies that look similar to the ones given in  Figure~\ref{fig:diefreq2}. If we look at the relative frequencies (i.e.\ the frequency divided by the total number of trials), we see that these converge to the theoretically predicted value of $\frac{1}{6}$ as our number of trials grows larger.

This fundamental finding is also called the ``Borel's law of large numbers''.

\begin{theorem}[Borel's law of large numbers]
	Let $\Omega$ be a sample space of some experiment and let $P$ be a probability mass function on $\Omega$. Furthermore let $N_n(E)$ be the number of occurrences of the event $E\subset \Omega$ when the experiment is repeated $n$ times. Then the following holds: 
    $$\frac{N_n(E)}{n}\rightarrow P(E) \quad \mathrm{as}\ {n\to\infty}.$$
\end{theorem}

Borel's law of large numbers states that if an experiment is repeated many times (where the trials have to be independent and done under identical conditions), then the relative frequency of the outcomes converge to their probability as assigned by the probability mass function. The theorem thus establishes the notion of probability as the long-run relative frequency of an events occurrence and thereby connects the theoretical side to the experimental side. Keep in mind though that we can never directly measure probabilities and although relative frequencies will converge to the probability values, they will usually not be exactly equal.


\subsection{Independence of Events and Conditional Probabilities}

A fundamental notion in probability theory is the idea of independence of events. Intuitively, we call two events independent if the occurrence of one does not affect the probability of occurrence of the other. Consider for example the events that it rains and the event that the current day of the week is Monday. These two are clearly independent, unless we lived in a world where there would be a correlation between the two, i.e.\ where the probability of rain would be different on Mondays compared to the other days of the week which is clearly not the case.  

Similarly, we establish the notion of independence of two events in the sense of probability theory as follows.

\begin{definition}[independent events]
	Let $A$ and $B$ be two events of some probability space $(\Omega,\Sigma,P)$. Then $A$ and $B$ are called \emph{independent} if and only if	
    \begin{equation}
    	P(A \cap B) = P(A)P(B).
    	\label{eqn:IndEvents}	
    \end{equation}
\end{definition}

\begin{figure}
\centering
	\includegraphics{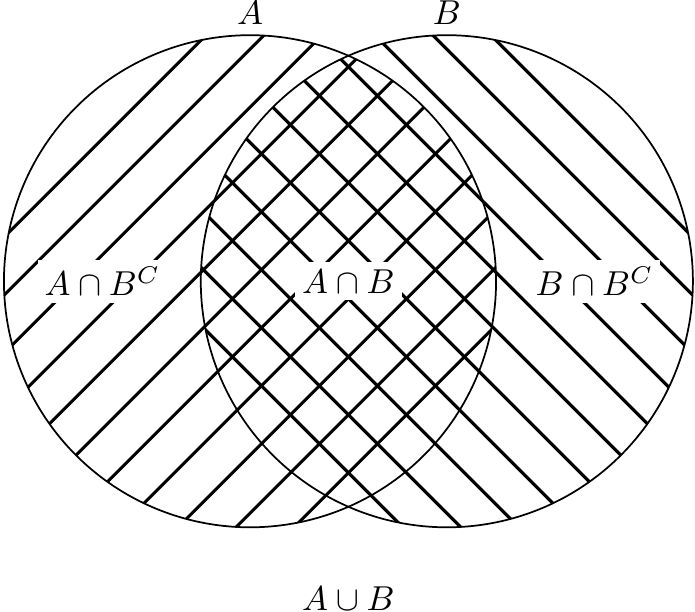}
\caption{Two events $A$ and $B$, their union $A\cup B$, their intersection $A\cap B$ (i.e.\ common occurrence in terms of probability) and their exclusive occurrences $A\cap B^C$ ($A$ and not $B$ occurs), $B\cap A^C$ ($B$ occurs and not $A$), where ${\cdot}^C$ denotes the complement in $A\cup B$.}
\label{fig:vennab}
\end{figure}

The term $P(A \cap B)$ is referred to \emph{joint probability} of $A$ and $B$, see Figure~\ref{fig:vennab}.

Another important concept is the notion of conditional probability, i.e.\ the probability of one event $A$ occurring, given the fact that another event $B$ occurred.

\begin{definition}[conditional probability]
	Given two events $A$ and $B$ of some probability space $(\Omega,\mathcal{F},P)$ with $P(B)>0$ we call
	
	\begin{equation*}
		P(A|B)=\frac{P(A\cap B)}{P(B)}
	\end{equation*}
	
	the \emph{conditional probability of $A$ given $B$}. 
\end{definition}

Note that for independent events $A$ and $B$, we have $P(A \cap B) = P(A)P(B)$ and thus $P(A|B)=P(A)$ and $P(B|A)=P(B)$. We can thus write 

\begin{equation*}
\begin{aligned} 
P(A \cap B) &= P(A)P(B),\\
\Leftrightarrow P(A) &= \frac{P(A \cap B)}{P(B)} = P(A\mid B),\\
\Leftrightarrow P(B) &= \frac{P(A \cap B)}{P(A)} = P(B\mid A), 
\end{aligned} 
\end{equation*}

and this means that the occurrence of $A$ does not affect the conditional probability of $B$ given $A$ (and vice versa). This exactly reflects the intuitive definition of independence that we gave in the first paragraph of this section. Note that we could have also used the conditional probabilities to define independence in the first place. None the less the definition of Equation~\ref{eqn:IndEvents} is preferred, as it is shorter, symmetrical in $A$ and $B$ and more general as the conditional probabilities above are not defined in the case where $P(A)=0$ or $P(B)=0$.

\subsection{Random Variables}
\label{ssec:randvar}

In many cases the sample spaces of random experiments are a lot more complicated than the ones of the toy examples we looked at so far. Think for example of measurements of membrane potentials of certain neurons, that we want to model mathematically, or the state of some complicated system, e.g.\ a network of neurons receiving some stimulus. 

Thus mathematicians came up with a way to tame the sample spaces by looking at the events indirectly, namely by first mapping the events to some better understood space, like the set of real numbers (or some higher dimensional real vector space) and then look at outcomes of the random experiment in the simplified space rather than in the complicated original space. Looking at spaces of numbers has many advantages: order relations exist (smaller, equal, larger), we can form averages and much more. This leads to the concept of random variables.

A (real) \emph{random variable} is a function that maps each outcome of a random experiment to some (real) number. Thus, a random variable can be thought of as a variable whose value is subject to variations due to chance. But keep in mind that a random variable is a mapping and not a variable in the usual sense.






Mathematically, a random variable is defined using what is called a \emph{measurable function}. A measurable function is nothing more than a map from one measurable space to another for which the pre-image of each measurable set is again measurable (with respect to the two different measures in the two measure spaces involved). So a measurable map is nothing more than a ``nice'' map respecting the structures of the spaces involved (take as an example for such maps the continuous functions over $\mathbb{R}$).

\begin{definition}[random variable]
    Let $(\Omega,\Sigma,P)$ be a probability space and $(\Omega',\Sigma')$ a measure space. A $(\Sigma,\Sigma')$-measurable function  $X\colon\Omega\to\Omega'$ is called \emph{$\Omega'$-valued random variable} (or just \emph{$\Omega'$-random variable}) on $\Omega$ .
\end{definition}

Commonly, a distinction between \emph{continuous random variables} and \emph{discrete random variables} is made, the former taking values on some continuum (in most cases $\mathbb{R}$) and the latter on a discrete set (in most cases  $\mathbb{Z}$).


A type of random variable that plays an important role in modeling is the the so called \emph{Bernoulli random variable} that only takes two distinct values $0$ with probability $p$ and $1$ with probability $1-p$ (i.e.\ it has a Bernoulli distribution as its underlying probability distribution). Spiking behavior of a neuron is often modeled that way, where $1$ stands for ``neuron spiked'' and $0$ for ``neuron didn't spike'' (in some interval of time). 

A real- or integer-valued random variable $X$ thus assigns a number $X(E)$ to every event $E\in \Sigma$. A value $X(E)$ corresponds to the occurrence of the event $E$ and is called a \emph{realization of X}. 
Thus, random variables allow for the change of space in which outcomes of probabilistic processes are considered. Instead of considering an outcome directly in some complicated space, we first project it to a simpler space using our mapping (the random variable $X$) and interpret its outcome in that simpler space.

In terms of measure theory, a random variable $X: (\Omega,\Sigma,P) \rightarrow (\Omega',\Sigma')$ (again, considered as a measurable mapping here) induces a probability measure $P_X$ on the measure space $(\Omega',\Sigma')$ via

\begin{equation*}
	P_X(S'):=P(X^{-1}(S')),
\end{equation*}

where again $X^{-1}(S')$ denotes the pre-image of $S'\in \Sigma'$. This also justifies the restriction of $X$ to be measurable: If it were not, such a construction would not be possible, but this is a technical detail. As a result, this makes $(\Omega',\Sigma',P_X)$ a probability space and we can think of the measure $P_X$ as the ``projection'' of the measure $P$ from $\Omega$ onto $\Omega'$ (via the measurable mapping $X$).

The measures $P$ and $P_X$ are probability densities for the probability distributions over $\Omega$ and $\Omega'$: They measure the likelihood of occurrence for each event ($P$) or value ($P_X$).

As a simple example of a random variable consider again the example of the  coin toss. Here, we have $\Omega = \{$heads,tails$\}$, $\mathcal{F} = \{\emptyset, \{$heads$\}, \{$tails$\}, \Omega \}$ and $P$ that assigns to both heads and tails the probability $\frac{1}{2}$ forming the probability space. Consider as a random variable $X: \Omega \rightarrow \Omega'$ with $\Omega'=\{0,1\}$ that maps $\Omega$ to $S$ such that $X($heads$)=0$ and $X($tails$)=1$. If we choose $\mathcal{F}'=\{\emptyset, \{0\}, \{1\}, \{0,1\}\}$ as a $\sigma$-algebra for $\Omega'$ this makes $M=(\Omega',\mathcal{F}')$ a measurable space and $X$ induces a measure $P'=P_X$ on $M$ with $P'(\{0\})=P'(\{1\})=\frac{1}{2}$. That makes $(\Omega',\mathcal{F}',P')$ a measure space and since $P'$ is normed it is a probability space.



\subsubsection*{Cumulative Distribution Function}

Using random variables that take on values of whole or the real numbers, the natural total ordering of elements in these spaces enables us to define the so called \emph{cumulative distribution function} (or \emph{cdf}) for a random variable.

\begin{definition}[cumulative distribution function]
	Let $X$ be a $\mathbb{R}$-valued or $\mathbb{Z}$-valued random variable on some probability space $(\Omega,\Sigma,P)$. Then the function
	
	\begin{equation*}
		F(x):=P(X\leq x)
	\end{equation*}
	
	is called the \emph{cumulative distribution function} of $X$. 
\end{definition}

The expression expression $P(X\leq x)$ evaluates to

\begin{equation*}
P(X\leq x)=\int_{\tau\leq x} P(X=\tau) \;\mathrm{d}\tau,
\end{equation*}

in the continuous case and to

\begin{equation*}
	P(X\leq x)=\sum_{k\leq x} P(X=k)
\end{equation*}

in the discrete case.

In that sense, the measure $P_X$ can be understood as the derivative of the cumulative distribution function $F$

\begin{equation*}
	P(x_1\leq X\leq x_2)=F(x_2)-F(x_1),  
\end{equation*}

and we also write $F(x)=\int_{\tau\leq x} P_X(\tau) \;\mathrm{d}\tau$ in the continuous case.

\subsubsection*{Independence of Random Variables}

The definition of independent events directly transfers to random variables: Two random variables $X, Y$ are called independent if the conditional probability distribution of $X$ ($Y$) given an observed value of $Y$ ($X$) does not differ from the probability distribution of $X$ ($Y$) alone.

\begin{definition}[independent random variables]
	Let $X,Y$ be two random variables. Then $X$ and $Y$ are called \emph{independent}, if the following holds for any observed values $x$ of $X$ and $y$ of $Y$:
	
	\begin{equation*}
		P(X|Y=y)=P(X) \quad \mathrm{and} \quad P(Y|X=x)=P(Y). 
	\end{equation*}
	
\end{definition}

This notion can be generalized to the case of three or more random variables naturally.   

\subsubsection*{Expectation and Variance}

Two very important concepts of random variables are the so called \emph{expectation value} (or just \emph{expectation}) and the \emph{variance}. The expectation of a random variable $X$ is the mean value of the random variable, where the weighting of the values corresponds to the probability density distribution. It thus tells us what value of $X$ we should expect ``on average'':

\begin{definition}[expectation value]
	Let $X$ be a $\mathbb{R}$- or $\mathbb{Z}$-valued random variable. Then its \emph{expectation value} (sometimes also denoted by $\mu$) is given by
	
	\begin{equation*}
		E[X]:=\int_{\mathbb{R}} x P_X(x)\; \mathrm{d}x=\int_{\mathbb{R}} x \; \mathrm{d}P_X,
	\end{equation*}

	for a real-valued random variable $X$ and by
	
	\begin{equation*}
		E[X]:=\sum_{x\in\mathbb{Z}} x P_X(x)
	\end{equation*}
	
	 if $X$ is $\mathbb{Z}$-valued. 
\end{definition}

Note that if confusion can be made as to which probability distribution the expectation value is taken, we will include the probability distribution to which the expectation value is taken in the index. Consider for example two random variables $X$ and $Y$ defined on the same base space but with different underlying probability distributions. In this case, we denote by $E_X[Y]$ the expectation value of $Y$ taken with respect to the probability distribution of $X$.

Let us now look an example. If we consider the throw of a fair die with $P(i)=\frac{1}{6}$ for each digit $i=1,\dots,6$ and take $X$ as the random variable that just assigns each digit its integer value $X(i)=i$, we get $E[X]=\frac{1}{6} (1+\dots+6)=3.5$.

Another important concept is the so-called \emph{variance} of a random variable. The variance is a measure for how far the values of the random variable are spread around its expected value. It is defined as follows.

\begin{definition}[variance]
	Let $X$ be a $\mathbb{R}$- or $\mathbb{Z}$-valued random variable. Then its \emph{variance} is given as
	
	\begin{equation*}
		\mathrm{var}[X]:=E[(E[X]-X)^2]=E[X^2]-(E[X])^2,
	\end{equation*}		
	
	sometimes also denoted as $\sigma^2$.	
\end{definition}

The variance is thus the expected squared distance of the values of the random variable to its expected value. 

Another commonly used measure is the so called \emph{standard deviation} $\sigma(X)=\sqrt{\mathrm{var}(X)}$, a measure for the average deviation of realizations of $X$ from the mean value.

Often one also talks about the expectation value as ``first order moment'' of the random variable, the variance as a ``second order moment''. Higher order moments can be constructed by iteration, but will not be of interest to us in the following.

	
	

	
	

Note again that the concepts of expectation and variance live on the theoretical side of the world, i.e.\ we cannot measure these quantities directly. The only thing that we can do is try to estimate them from a set of measurements (i.e.\ realizations of the involved random variables), see Figure~\ref{fig:theoryworld}. The statistical discipline of estimation theory deals with question regarding the estimation of theoretical quantities from real data. We will talk about estimation in more detail in Section~\ref{sec:estimation} and just give two examples here.

For estimating the expected value we can use what is called the \emph{sample mean}.

\begin{definition}[sample mean]
	Let $X$ be a $\mathbb{R}$- or $\mathbb{Z}$-valued random variable with $n$ realizations $x_1,\dots,x_n$. Then the \emph{sample mean $\hat{\mu}$} of the realizations is given as 
	\begin{equation*}
		\hat{\mu}(x_1,\dots,x_n):=\frac{1}{n} \sum_{i=1}^n x_i
	\end{equation*}
\end{definition}

As we will see below, this sample mean provides a good estimation of the expected value if the number $n$ of samples is large enough. Similarly, we can estimate the variance as follows.

\begin{definition}[sample variance]
	Let $X$ be a $\mathbb{R}$- or $\mathbb{Z}$-valued random variable with $n$ realizations $x_1,\dots,x_n$. Then the \emph{population variance $\hat{\sigma}$} of the realizations is given as 
	\begin{equation*}
		\hat{\sigma}^2(x_1,\dots,x_n):=\frac{1}{n} \sum_{i=1}^n (x_i-\hat{\mu}(x_1,\dots,x_n))^2,
	\end{equation*}
	where $\hat{\mu}$ denotes the sample mean.
\end{definition}

Before going on let us calculate some examples of expectations and variances of random variables. Take the coin toss example from above. Here, the expected value of $X$ is $E[X]=\frac{1}{2}\cdot 0 + \frac{1}{2}\cdot 1 = \frac{1}{2}$, the variance $\operatorname{var}(X)=E[(E[X]-X)^2]=\frac{1}{2}\cdot (0-\frac{1}{2})^2 + \frac{1}{2}\cdot (1-\frac{1}{2})^2 = \frac{1}{4}$. For the example of the dice roll (where the random variable $X$ takes the value of the number thrown) we get $E[X] = \frac{1 + 2 + 3 + 4 + 5 + 6}{6} = \frac{7}{2} = 3.5$ and $\operatorname{var}(X)=E[X^2]-(E[X])^2=\frac{91}{6}- \frac{49}{4}=\frac{35}{12}\approx 2.92$.

\subsubsection*{Relations between the expectation value and the variance}

Studying relations between the expected value and the variance of a random variable can be helpful in order to get a clue on the shape of the underlying probability distribution. The quantities we will discuss below are usually only used in conjunction with positive statistics, such as count data or the time between events, but can be extended to the general case without greater problems (if this is needed).

A first fundamental quantity that can be derived as a relation between the expected value and the variance is the so called signal to noise ratio (remember that we can interpret a sequence of realizations of a random variable as a ``signal'').  

 \begin{definition}[signal to noise ratio]
 	Let $X$ be a non-negative real valued random variable with expected value $\mu\geq 0$, variance $\sigma^2\geq 0$ and standard deviation $\sigma$. Then its \emph{signal to noise ratio} (SNR) is given by

 	\begin{equation*}
 	\operatorname{SNR}(X)=\frac{\mu}{\sigma}.
 	\end{equation*}
	
\end{definition}

This dimensionless quantity has wide applications in physics and signal processing. Note that the signal to noise ratio approaches infinity as $\sigma$ approaches $0$ and that is defined to be infinity for $\sigma=0$. 

Being defined via the the expectation and variance, the signal to noise ratio again is a theoretical quantity that can only be computed if the full probability distribution of $X$ is known. As this usually is not the case, we have to resort on estimating $\operatorname{SNR}(X)$ via the so called \emph{population SNR} which we calculate for a given population (our samples) using the ratio of the sample standard mean $\bar{x}$, to the sample standard deviation $s$ to obtain

\begin{equation*}
    \widehat{\operatorname{snr}} = \frac{\bar{x}}{s}.
\end{equation*}

Keep in mind though that this estimation is \emph{biased}. This means that it tends to yield a value shifted with respect to the real value, in this case a value higher than $\operatorname{SNR}(X)$. This especially applies to cases with fewer samples. This means that $ \widehat{\operatorname{snr}}$ can be used for an estimation of an upper bound for $\operatorname{SNR}(X)$.

Another fundamental property defined via expectation and variance of a random variable are the so called \emph{index of dispersion} and the \emph{coefficient of variation}.

 \begin{definition}[index of dispersion, coefficient of variation]
 	Let $X$ be a a non-negative real valued random variable with expected value $\mu\geq 0$, variance $\sigma^2\geq 0$ and standard deviation $\sigma$. Then its \emph{index of dispersion} is given by
	
 	\begin{equation*}
 		D:=\frac{\sigma^2}{\mu}.
 	\end{equation*}

 	The \emph{coefficient of variation} (CV) is given by
		
 	\begin{equation*}
 	\operatorname{CV}(X):=\frac{\sigma}{\mu}.
 	\end{equation*}
 \end{definition}

We will only discuss the coefficient of variation in more detail in the following, but the same also holds for the index of dispersion.

Again, the coefficient of variation is a theoretical quantity that can be estimated via the \emph{population CV} analogously to the population SNR. Note that this estimator also is positively biased, i.e.\ it usually overestimates the value of $\operatorname{CV}(X)$.

There are some other noteworthy properties of the coefficient of variation. First of all, it is a dimensionless number and independent of the unit of the (sample) data, which is a desirable property, especially when comparing different data sets. This makes the CV a popular and often used quantity in statistics which is better suited than e.g.\ the standard deviation when comparing different data sets. But there are certain drawbacks to the quotient construction, too: The CV becomes very sensitive to fluctuations in the mean close to $\mu=0$ and it is undefined (or infinity) for a mean of $\mu=0$.  


Why do these quantities matter? They will allow us to distinguish certain families of probability distributions (as discussed in Section~\ref{sec:paramprobdistr}) --- for example a Poisson-distributed random variable has a coefficient of variation of $1$ and this is a necessary condition for the random variable to be Poisson-distributed.

\begin{figure}
\centering
 	\includegraphics{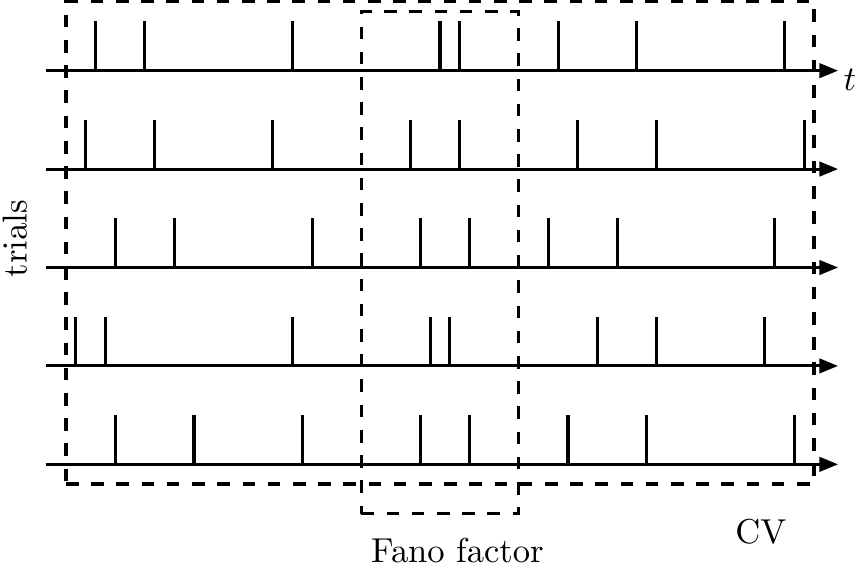}
\caption{Coefficient of variation versus Fano factor of a certain hypothesized set of trials of recordings of neuronal spiking activity of a single neuron (trials shown top to bottom): the Fano factor is the time-windowed version of the coefficient of variation (CV).}
\label{fig:cvfano}
\end{figure}





\subsection{Laws of Large Numbers}

The laws of large numbers (there exist two versions as we will see below) state that the sample average of a set of realizations of a random variable ``almost certainly'' converges the the random variable's expected value when the number of realizations grows to infinity.

\begin{theorem}[law of large numbers]
Let $X_1, X_2, \dots$ be an infinite sequence of independent, identically distributed random variables with expected values $E(X_1) = E(X_2) = \dots = \mu$. Let $\overline{X}_n=\frac{1}{n}(X_1+\cdots+X_n)$ be the sample average.  

\begin{enumerate}[(i)]
	\item \textbf{Weak law of large numbers}. The sample average converges in probability towards the expected value, i.e.\ for any $\varepsilon>0$ 
		\begin{equation*}
		\lim_{n\to\infty} P \left(\,|\overline{X}_n-\mu| > \varepsilon\,\right) = 0. 
		\end{equation*}
		This is sometimes also expressed as
		\begin{equation*}
		\overline{X}_n\ \xrightarrow{P}\ \mu \qquad\textrm{when}\ n \to \infty. 
		\end{equation*}
		
	\item \textbf{Strong law of large numbers}. The sample average converges almost surely towards the expected value, i.e.
		\begin{equation*}
			P \left( \lim_{n\to\infty}\overline{X}_n = \mu \right) = 1. 
		\end{equation*}
		This is sometimes also expressed as
		\begin{equation*}
		\overline{X}_n\ \xrightarrow{a.s.}\ \mu \qquad\textrm{when}\ n \to \infty. 
		\end{equation*}
		
\end{enumerate}

\end{theorem}

The weak version of the law states that the sample average $\overline{X}_n$ is likely to be close to $\mu$ for some large value of $n$. But this does not exclude the possibility of $|\overline{X}_n -\mu| > \varepsilon$ occurring an infinite number of times.

The strong law says that this ``almost surely'' will not be the case: With probability $1$, the inequality $|\overline{X}_n -\mu| < \varepsilon$ holds for all $\varepsilon>0$ and all large enough $n$.

\subsection{Some Parametrized Probability Distributions}
\label{sec:paramprobdistr}

Certain probability distributions often occur naturally when looking at typical random experiments. In the course of history, these were thus put (mathematicians like doing such things) into families or classes and the members of one class are distinguished by a set of parameters (a parameter is just a number than can be chosen freely in some specified range). To specify a certain probability distribution we simply have to specify in which class it lies and which parameter values it exhibits, which is more convenient than specifying the probability distribution explicitly every time. This also allows proving (and reusing) results for whole classes of probability distributions and, facilitates communication with other scientists. 


Note that we will only give a concise version of the most important distributions relevant in neuroscientific applications here and point the reader to \cite{Hoel71IntProbTheo, Shiryayev84Prob, Klenke08ProbTheo} for a more in-depth treatment of the subject.

The \emph{normal distribution} $N(\mu,\sigma)$ is a family of continuous probability distributions parametrized by two real-valued parameters $\mu\in \mathbb{R}$ and $\sigma^2\in \mathbb{R}^{+}$, called \emph{mean} and \emph{variance}. Its probability density function is given as

\begin{equation*}
\begin{aligned}
	f(x;\mu,\sigma)\,:\: \mathbb{R}&\rightarrow& \mathbb{R}_{0}^{+}\\
	x&\mapsto& \frac{1}{\sigma \sqrt{2\pi}} e^{-\frac{1}{2}(\frac{x-\mu}{\sigma})^2}.
\end{aligned}
\end{equation*} 

\begin{figure}
\centering
\subfigure[]{%
\label{fig:distnormalpdf}%
\includegraphics{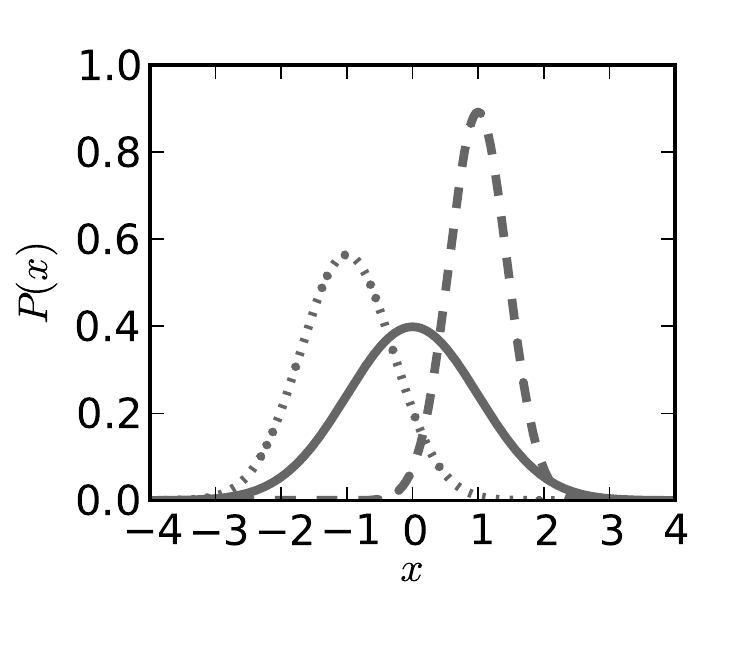}
}
\subfigure[]{%
\label{fig:distnormalcdf}%
\includegraphics{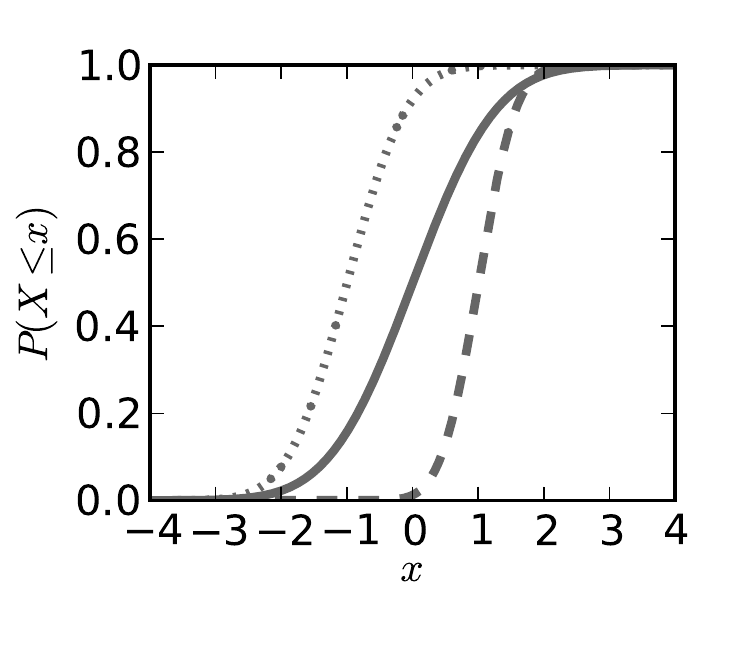}
}
\caption{Normal distribution: probability density function \subref{fig:distnormalpdf} and cumulative density function \subref{fig:distnormalcdf} for selected parameter values of $\mu$ and $\sigma$. Solid line: $\mu=0$, $\sigma^2=1$, dashed line: $\mu=1$, $\sigma^2=0.2$, dotted line: $\mu=-1$, $\sigma^2=0.5$.}
\label{fig:distnormal}
\end{figure}

The family is closed under linear combinations, i.e.\ linear combinations of normally distributed random variables are again normally distributed. It is the most important and often used probability distribution in probability theory and statistics as many other probability distributions can be approximated by a normal distribution when the sample size is large enough (this fact is called the \emph{central limit theorem}). See Figure~\ref{fig:distnormal} for examples of the pdf and cdf for normally-distributed random variables.

The \emph{Bernoulli probability distribution} $\operatorname{Ber}(p)$ describes the two possible outcomes of a Bernoulli experiment with the probability of success and failure being $p$ and $1-p$, respectively. It is thus a discrete probability distribution on two elements and it is parametrized by one parameter $p\in [0,1]\subset \mathbb{R}$. Its probability mass function is given by the two values $P(\mathrm{success})=p$ and $P(\mathrm{success})=1-p$.

The \emph{binomial probability distribution} $\operatorname{B}(n,p)$ is a discrete probability distribution  parametrized by two parameters $n\in \mathbb{N}$ and $p\in [0,1]\subset \mathbb{R}$. Its probability mass function is

\begin{equation}
	f(k;n,p)=\binom{n}{k} p^k (1-p)^{n-k},
\end{equation}

and it can be thought of as a model for the probability of $k$ successful outcomes in a trial with $n$ independent Bernoulli experiments, each having success probability $p$.


The \emph{Poisson distribution} $\operatorname{Poiss}(\lambda)$ is a family of discrete probability distributions parametrized by one real parameter $\lambda\in \mathbb{R}^{+}$. Its probability mass function is given by

\begin{equation*}
\begin{aligned}
	f(k;\lambda)\,:\: \mathbb{N}&\rightarrow&\mathbb{R}_{0}^{+}\\
	k&\mapsto& \frac{\lambda^k e^{-\lambda}}{k!}.
\end{aligned}
\end{equation*}

The Poisson distribution plays an important role in the modeling of neuroscience data. This is the case because the firing statistics of cortical neurons (and also other kinds of neurons) can often be well fit by a Poisson process, where $\lambda$ is considered the mean firing rate of a given neuron, see \cite{Softky1993, Compte2003, Nawrot2008}.

This fact comes at no surprise if we invest some thought. The Poisson distribution can be seen as a special case of the binomial distribution. 
A theorem known as Poisson limit theorem (sometimes also called ``law of rare events'') now tells us that in the limit $p\rightarrow 0$ and $n\rightarrow \infty$ the binomial distribution converges to the Poisson distribution with $\lambda=np$. Consider for example the spiking activity of our neuron that we could model via a Binomial distribution. We discretize time and consider time bins of say $2$ ms and assume a mean firing rate of the neuron denoted by $\lambda$ (measured in Hertz). Clearly, in most time bins the neuron does not spike (corresponding to a small value of $p$) and the number of bins is large (corresponding to a large $n$). The Poisson limit theorem tells us that in this case the probability distribution concerning spike emission is well matched by a Poisson distribution.     

\begin{figure}
\centering
\subfigure[]{%
\label{fig:distpoisspmf}%
\includegraphics{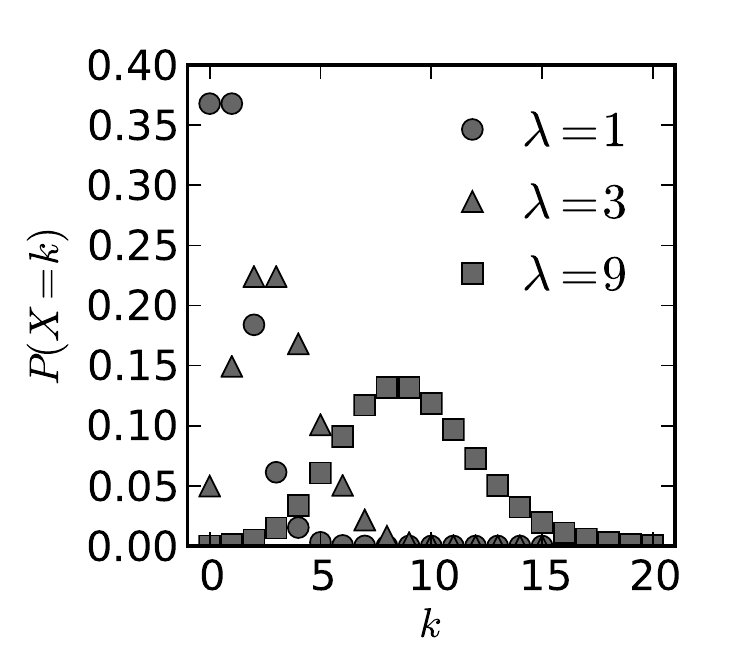}
}
\subfigure[]{%
\label{fig:distpoisscdf}%
\includegraphics{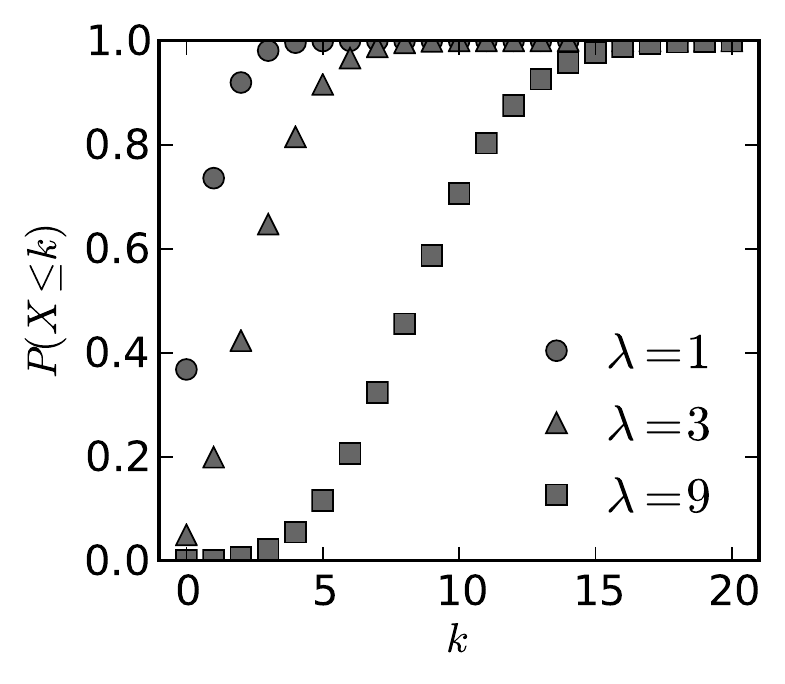}
}
\caption{Poisson distribution: probability mass function \subref{fig:distpoisspmf} and cumulative density function \subref{fig:distpoisscdf} for selected parameter values of $\lambda$.}
\label{fig:distpoiss}
\end{figure}

See Figure~\ref{fig:distpoiss} for examples of the pmf and cdf for Poisson-distributed random variables for a selection of parameters $\lambda$.

\begin{figure}
\centering
\subfigure[]{%
\label{fig:distbinpmf}%
\includegraphics{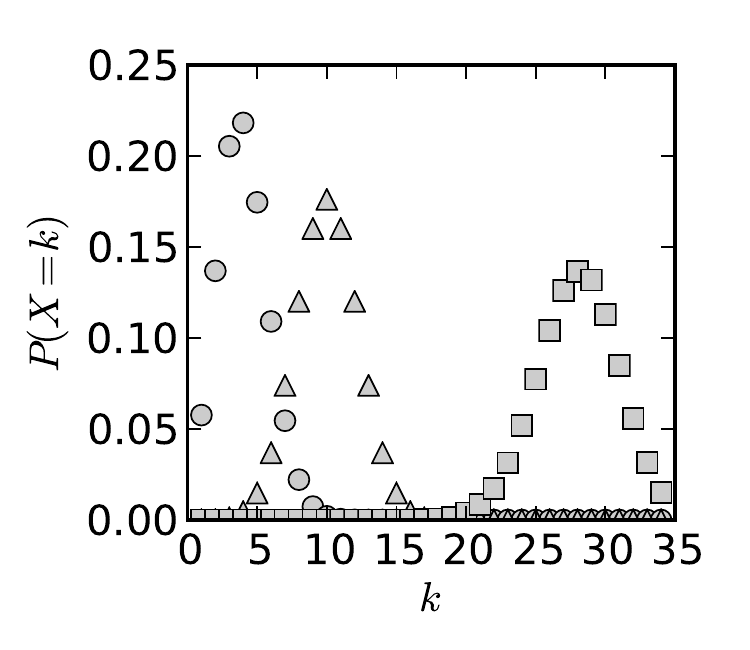}
}
\subfigure[]{%
\label{fig:distbincdf}%
\includegraphics{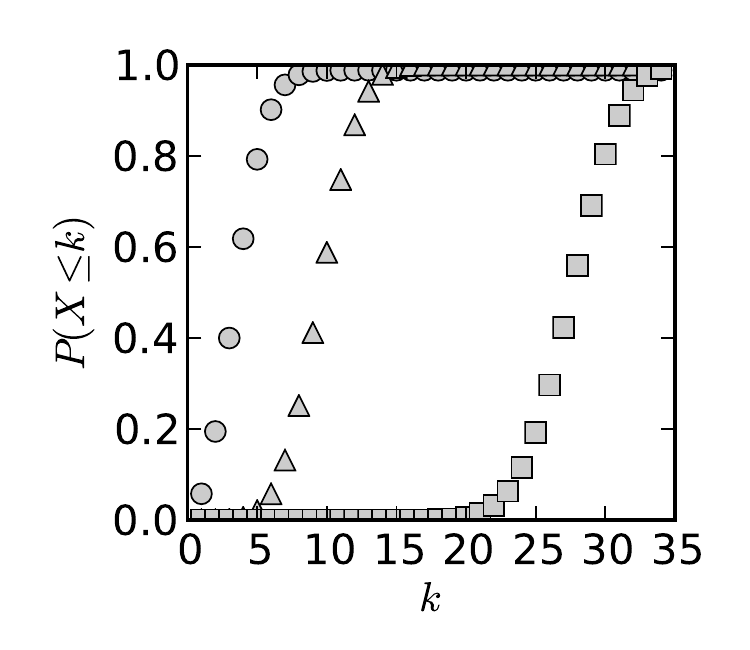}
}
\caption{Binomial distribution: probability mass function \subref{fig:distbinpmf} and cumulative density function \subref{fig:distbincdf} for selected parameter values of $p$ and $n$. Circle: $p=0.2$, $n=20$, triangle: $p=0.5$, $n=20$, square: $p=0.7$, $n=40$.}
\label{fig:distbin}
\end{figure}

The so called \emph{exponential distribution} $\operatorname{Exp}(\lambda)$ is a continuous probability distribution parametrized by one real parameter $\lambda\in \mathbb{R}^{+}$. Its probability density function is given by

\begin{equation*}
\begin{aligned}
	f(x;\lambda)\,:\: \mathbb{R}&\rightarrow& \mathbb{R}_{0}^{+}\\
	x&\mapsto& \left\{ \begin{array}{ll}\lambda e^{-\lambda x} &\mathrm{for}\ x> 0\\ 0  &\mathrm{for}\ x\leq 0\end{array} \right..
\end{aligned}
\end{equation*}

The exponential distribution with parameter $\lambda$ can be interpreted as the probability distribution describing the time between two events in a Poisson process with parameter $\lambda$, see the next section.

\begin{figure}
\centering
\subfigure[]{%
\label{fig:distexppdf}%
\includegraphics{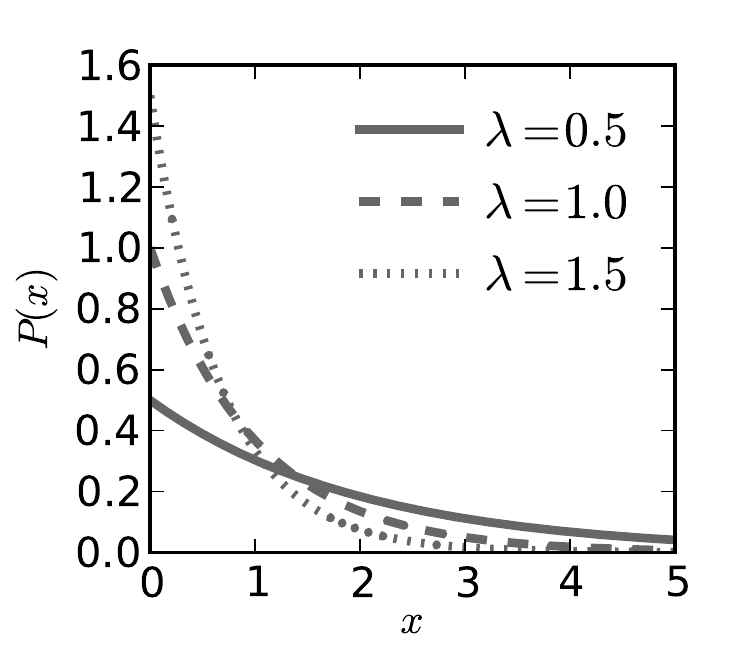}
}
\subfigure[]{%
\label{fig:distexpcdf}%
\includegraphics{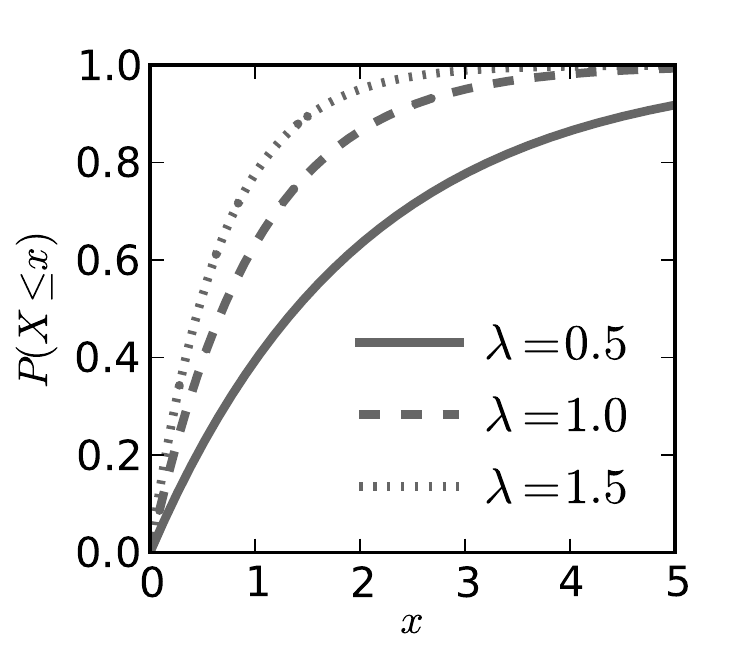}
}
\caption{Exponential distribution: probability density function \subref{fig:distexppdf} and cumulative density function \subref{fig:distexpcdf} for selected parameter values of $\lambda$.}
\label{fig:distexp}
\end{figure}

See Figure~\ref{fig:distexp} for examples of the pdf and cdf for exponentially-distributed random variables for a selection of parameters $\lambda$.

We want to conclude our view on families on probability distributions at this point and point the interested reader to \cite{Hoel71IntProbTheo, Shiryayev84Prob, Klenke08ProbTheo} regarding further examples and details of families of probability distributions.

\subsection{Stochastic Processes}
\label{sec:stochasticproc}

A \emph{stochastic process} (sometimes also called \emph{random process}) is a collection of random variables indexed by a totally ordered set, which is usually taken as time. Stochastic processes are commonly used to model the evolution of some random variable over time. We will only look at discrete-time processes in the following, i.e.\ stochastic processes that are indexed by a discrete set. The extension to the continuous case is straightforward, see \cite{Brzezniak1999} for an introduction to the subject.

Mathematically, a stochastic process is defined as follows.

\begin{definition}
Let $(\Omega, \mathcal{F}, P)$ be a probability space and let $(S,\mathcal{S})$ be a measure space. Let furthermore $X_t: \mathcal{F} \rightarrow \mathcal{S}$ be a set of random variables, where $t\in T$. Then an \emph{$S$-valued stochastic process $\mathcal{P}$} is given by   

\begin{equation*}
    \mathcal{P}:=\{ X_t : t \in T \},
\end{equation*}

where $T$ is some totally ordered set, commonly interpreted as time. The space $S$ is referred to as the \emph{sample space of the process $\mathcal{P}$}.  
\end{definition}

If the distribution underlying the random variables $X_t$ does not vary over time, the process is called \emph{homogeneous}, in the case where the probability distributions $P_{X_t}$ depend on the time $t$, it is called \emph{inhomogeneous}.


A special kind and well-studied type of stochastic process is the so called \emph{Markov process}. A discrete Markov process of order $k\in \mathbb{N}$ is a inhomogeneous stochastic process subject to the restriction that for any time $t=0,1,\dots$, the probability distribution underlying $X_t$ only depends on the preceding $k$ probability distributions of $X_{t-1},\dots,X_{t-k}$, i.e. that for any $t$ and any set of realizations $x_i$ of $X_i$ ($0\leq i\leq t$) we have

\begin{equation*}
P(X_t=x_t|X_{t-1}=x_{t-1}, \dots, X_{t-k}=x_{t-k})=
P(X_t=x_t|X_{t-1}=x_{t-1}, \dots, X_0=x_0). 
\end{equation*}


						




\begin{figure}
\centering
\subfigure[]{%
\label{fig:procpoissstep}%
\includegraphics{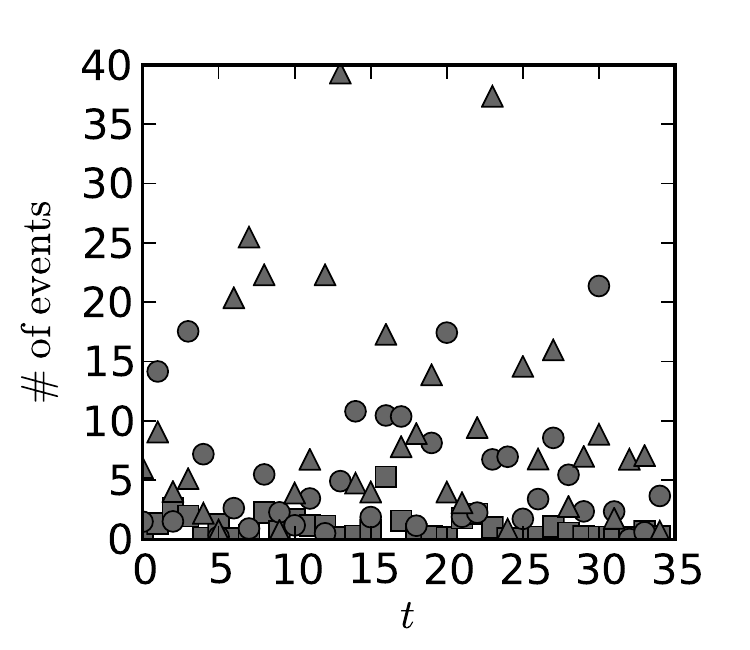}
}
\subfigure[]{%
\label{fig:procpoissaccum}%
\includegraphics{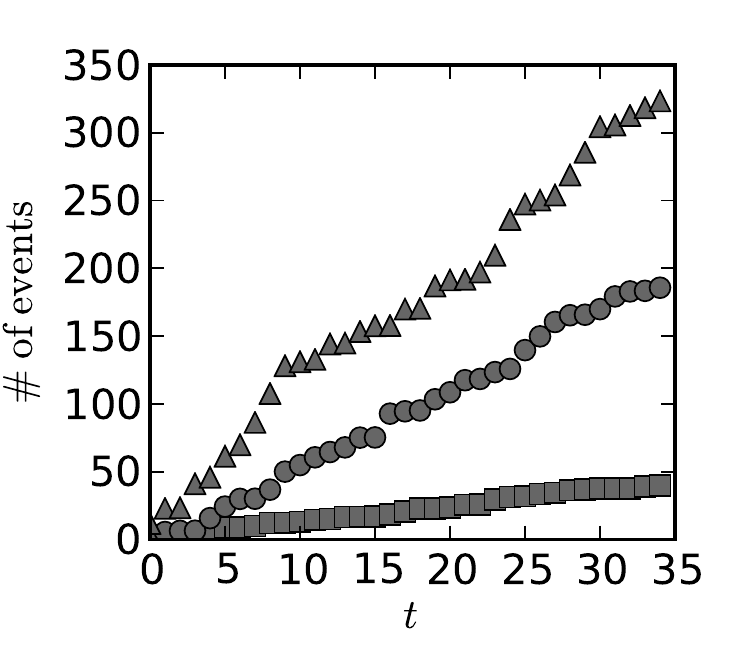}
}
\caption{Examples of the number of events in one time window of size $\Delta t=1$ \subref{fig:procpoissstep} and the number of accumulated events since $t=0$ \subref{fig:procpoissaccum} for Poisson processes with rates $\lambda=1$ (circle), $\lambda=5$ (triangle) and $\lambda=10$ (square).}
\label{fig:procpoiss}
\end{figure}

Another process often considered in neuroscientific applications if the \emph{Poisson process}. It is a discrete-time stochastic process $\mathcal{P}$ for which the random variables are Poisson-distributed with some parameter $\lambda(t)$ (in the inhomogeneous case, for the homogeneous case we have $\lambda(t)=\lambda=$constant). As can be shown, the time delay between each pair of consecutive events of a Poisson process is exponentially distributed. See Figure~\ref{fig:procpoiss} for examples of the number of instantaneous (occurring during one time slice) and the number of cumulated events (over all preceding time slices) of Poisson processes for a selection of parameters $\lambda$.  

Poisson processes have proven to be a good model for many natural as well as man-made processes such as radioactive decay, telephone calls and queues, and also for modeling neural data.  An influential paper in the neurosciences was \cite{Colquhoun85FastEvSingleChanCurr}, showing the random nature of the closing and opening of single ion channels in certain neurons. Using as model a Poisson process with the right parameter provides a good fit to the measured data here. 

Another prominent example of neuroscientific models employing a Poisson process is the commonly used model for the sparse and highly irregular firing patterns of cortical neurons in vivo \cite{Softky1993, Compte2003, Nawrot2008}. The firing patterns of such cells are usually modeled using inhomogeneous Poisson processes (with $\lambda(t)$ modeling the average firing rate of a cell). 


\begin{figure}
\centering
	\includegraphics{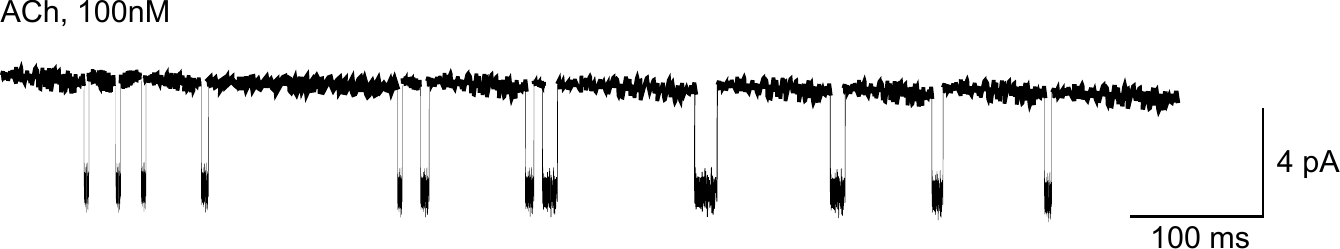}
\caption{Random opening and closing of ion channels. Modified from \cite{Colquhoun85FastEvSingleChanCurr}, Figure~12.}
\label{fig:ionchan1}
\end{figure}






\section{Information Theory}
\label{sec:informationentropy}

\begin{figure}
\centering
	\includegraphics{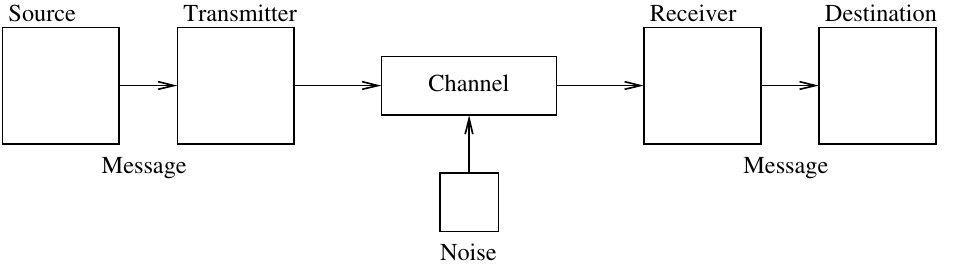}
\caption{The setting of Shannon's information theory: information is transferred from a source to a destination via a message that is first encoded, and then subsequently sent over a noisy channel to be decoded by the receiver.}
\label{fig:shannonchannel}
\end{figure}

Information theory was introduced by Shannon \cite{Shannon1948} as a mathematically rigid theory to describe the process of transmission of information over some channel of communication. His goal was quantitatively measure the ``information content'' of a ``message'' sent over some ``channel'', see Figure~\ref{fig:shannonchannel}. In what follows we will not go into detail regarding all aspects of Shannon's theory, but we will mainly focus on his idea of measuring ``information content'' of a message. For a more in-depth treatment of the subject, the interested reader is pointed to the excellent book \cite{Cover1991}.

The central elements of Shannon's theory are depicted in Figure~\ref{fig:shannonchannel}. In the standard setting considered in information theory, an \emph{information source} produces \emph{messages} that are subsequently encoded using \emph{symbols} from an \emph{alphabet} and sent over a noisy \emph{channel} to be received by a \emph{receiver} that decodes the message and attempts to reconstruct the original message.

A communication channel (or just channel) in Shannon's model transmits the encoded message from the sender to the receiver. Due to noise present in the channel the receiver does not receive the original message dispatched by the sender but rather some noisy version of it. 

The whole theory is set in the field of probability theory (hence our introduction to the concepts in the last section) and in this context, the messages emitted by the source are modeled as a random variable $X$ with some underlying probability distribution $P_X$. For each message $x$ (a realization of $X$), the receiver sees a corrupted version $y$ of $x$ and this fact is modeled by interpreting the received messages as realizations of a random variable $Y$ with some probability distribution $P_Y$ (that depends both on $P_X$ and the channel properties). The transmission characteristics of the channel itself are characterized by the stochastic correspondence of the signals transmitted by the sender to the ones received by the receiver, i.e.\ by modeling the channel as a conditional probability distribution $P_{Y|X}$.

Being based upon probability theory, keep in mind that the all the information-theoretic quantities that we will look at in the following such as ``entropy'' or ``mutual information'' are just properties of the random variables involved, i.e.\ properties of the probability distributions underlying these random variables.

Information-theoretic analyses have proven to be a valuable tool in many areas of science such as physics, biology, chemistry, finance and linguistics and generally in the study of complex systems \cite{Prokopenko2009, Lizier2010}. We will have a look at applications in the neurosciences in Section~\ref{sec:neuralsystems}.

Note that a vast number of works was published in the field of information theory and its applications since its first presentation in the 1950s. We will focus on the core concepts in the following and point the reader to \cite{Cover1991} for a more in-depth treatment of the subject. 

In the following we will start by looking at a notion of information and using this proceed to define \emph{entropy} (sometimes also called \emph{Shannon entropy}), a core concept in information theory. As all further information-theoretic concepts are based on the idea of entropy, it is of vital importance to understand this concept well. We will then look at mutual information, the information shared by two or more random variables. Furthermore, we will look at a measure of distance for probability distributions called Kullback-Leibler Divergence and give an interpretation of mutual information in terms of Kullback-Leibler Divergence. After a quick look at the multivariate case of mutual information between more than two variables and the relation between mutual information and channel capacity we will then proceed to a information-theoretic measure called transfer entropy. Transfer entropy is based on mutual information but in contrast to mutual information is of directed nature.

\subsection{A Notion of Information}
\label{ssec:information}

Before defining entropy, let us try to give an axiomatic definition of the concept of ``information'' \cite{Csiszar2008}. The entropy of a random variable will then be nothing more that the expected (i.e.\ average) amount of information contained in a realization of that random variable.

We want to consider a probabilistic model in what follows, i.e. we have a set of events, each occurring with a given probability. The goal is to assess how informative the occurrence of a given event is. What would we intuitively expect from a measure of information $h$ that maps the set of the events to the set of non-negative real number, i.e.\ when we restrict $h$ to be a non-negative real number?

First of all, it should certainly be additive for independent events and sub-additive for non-independent events. This is easily justified: If you read two newspaper articles about totally unrelated subjects, the total amount of information you obtain consists of both the information in the first and the second article. When you read articles about related subjects, they often have some common information. 

Furthermore, events that occur regularly and unsurprisingly are not considered informative and the more seldom or surprising an event occurs, the more informative it is. Think about an article about your favorite sports team winning a match that usually wins all matches. You will consider this not very informative. But when the local newspaper reports about an earthquake with its epicenter in the part of town where you live, this will certainly be informative to you (unless you were at home during the time the earthquake happened), assuming that earthquakes do not occur on a regular basis where you live. 

We thus have the following axioms for the information content $h$ of an event, where we look at the information content of events contained in some probability space $(\Omega,\Sigma,P)$.

\begin{enumerate}[(i)]
\item $h$ is non-negative: $h: \Sigma \rightarrow \mathbb{R}^{+}$.
\item $h$ is sub-additive: For any two messages $\omega_1,\omega_2\in \Sigma$ we have $h(\omega_1\cap \omega_2)\leq h(\omega_1)+h(\omega_2)$, where equality holds if and only if $\omega_1$ and $\omega_2$ are independent.
\item $h$ is continuous and monotonic with respect to the probability measure $P$.
\item Events with probability $1$ are not informative: $h(\omega)=0$ for $\omega\in\Sigma$ with $P(\omega)=1$.
\end{enumerate}

Now calculus tells us (this is not hard to show --- you paid attention in the mathematics class at school, didn't you?) that these four requirements leave only one possible function that fulfills all these requirements: the logarithm. This leads us to the following natural definition.

\begin{figure}
\centering
	\includegraphics{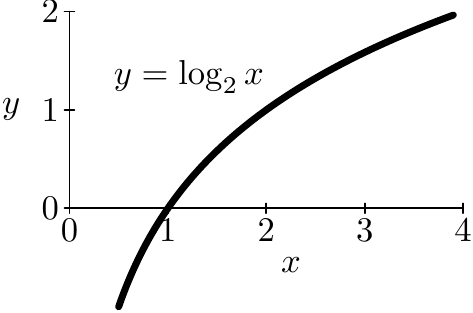}
\caption{The logarithm to the basis of $2$.}
\label{fig:log2}
\end{figure}

\begin{definition}[Information]
	Let $(\Omega,\Sigma,P)$ be a probability space. Then the information $h$ of an event $\sigma\in\Sigma$ is defined as
	
	\begin{equation*}
		h(\sigma):=h(P(\sigma))=-\log_{b}(P(\sigma)),
	\end{equation*}

	where $b$ denotes the basis of the logarithm.
\end{definition}

For the basis of the logarithm, usually $b=2$ or $b=e$ is chosen, fixing the unit of $h$ as ``bit'' or ``nat'', respectively. We resort to using $b=2$ for the rest of this chapter and write $\log$ for the logarithm to the basis of two. The natural logarithm will be denoted by $\ln$. 

Note that the information content in our definition only depends on the probability of the occurrence of the event and not the event itself. It is thus a property of the probability distribution $P$. 

Let us give some examples in order to illustrate this idea of information content.

Consider a toss of a fair coin, where the possible outcomes are heads (H) or tails (T), each occurring with probability $\frac{1}{2}$. What is the information contained in a coin toss? As the information solely depends on the probability, we have $h(H)=h(T)$, which comes at no surprise. Furthermore we have $h(H)=h(T)=-\log \frac{1}{2} =-(\log(1)-log_2(2))=\log 2 = 1$ bit, when we apply the fundamental logarithmic identity $\log(a\cdot b)=\log(a)+\log(b)$. Thus one toss of a fair coin gives us one bit of information. This fact also lets us explain the unit attached to $h$. If measured in bit (i.e.\ with $b=2$), this is the amount of bits needed to store that information. For the toss of a coin we need one bit, assigning each outcome to either $0$ or $1$.

Repeating the same game for the roll of a fair die where each digit has probability $\frac{1}{6}$, we again have the same amount of information for each digit $E\in\{1,\dots,6\}$, namely $h(E)=\log(6)\approx 2.58$ bit. This means that in this case we need $3$ bits to store the information associated to each outcome, namely the number shown.

Looking at the two examples above, we can give another (hopefully intuitive) characterization of the term information content: It is the minimal number of yes-no-questions that we have to ask until we know which event occurred, assuming that we have a knowledge of the underlying probability distribution. Consider the example of the coin toss above. We have to ask exactly one question and we know the outcome (``Was it heads?'', ``Was it tails?''). 

Things get more interesting when we look at the case of the die throw. Here several question asking strategies are possible and you can freely choose your favorite -- we will give one example below.

Say a digit $d$ was thrown. The first question could be ``Was the digit less or equal to 3?'' (other strategies ``Was the digit greater or equal to 3?'', ``Was the digit even?'', ``Was the digit odd?''). We then go on depending on the answer and cut off at least half of the remaining probability mass in each step, leaving us with a single possibility after at most $3$ steps. From the information content we know that on average we have to ask $2.58$ times on average.

The two examples above were both cases with uniform probability distributions but in principle the same applies to arbitrary probability distributions.

\subsection{Entropy as Expected Information Content}
\label{ssec:entropy}

The term entropy is at the heart of Shannon's information theory \cite{Shannon1948}. Using the notion of the information as discussed in Section~\ref{ssec:information}, we can readily define the entropy of a discrete random variable as its expected information. 

\begin{definition}[entropy]
	Let $X$ be a random variable on some probability space $(\Omega,\Sigma,P)$ with values in the integer or the real numbers. Then its \emph{entropy}\footnote{Shannon chose the letter $H$ for denoting entropy after Boltzmann's $H$-theorem in classical statistical mechanics.} (sometimes also called \emph{Shannon entropy} or \emph{self-information}) $H(X)$ is defined as the expected amount of information of $X$,
	
	\begin{equation}
		H(X):=E[h(X)].
		\label{eq:defentropy}
	\end{equation}
	
\end{definition}

If $X$ is a random variable that takes integer values (i.e.\ a discrete random variable), Equation~\ref{eq:defentropy} evaluates to 

\begin{equation*}
	H(X)=\sum_{x\in \mathbb{Z}} P(X=x) h(P(X=x))=-\sum_{x\in \mathbb{Z}} P(X=x) \log(P(X=x)),
\end{equation*}

in the case of a real-valued, continuous random variable we get

\begin{equation*}
H(X)=\int_{\mathbb{R}} P(X=x) h(P(X=x))\; \mathrm{d}x
\end{equation*}

and the resulting quantities is called \emph{differential entropy} \cite{Cover1991}.

As the information content is a function solely dependent on the probability of the events one also speaks of the entropy of a probability distribution. 

Looking at the definition in Equation~\ref{eq:defentropy}, we see that entropy is a measure for the average amount of information that we expect to obtain when looking at realizations of a given random variable $X$. An equivalent characterization would be to interpret it as the average information one is missing when one would not know the value of the random variable (i.e.\ its realization) and a third one would be to interpret it as the average reduction of uncertainty about the possible values of a random variable having observed one or more realizations.

Akin to the information content $h$, entropy $H$ is a dimensionless number and usually measured in bits (i.e.\ the expected number of binary digits needed to store the information) by taking a logarithm to the base of $2$.

Shannon entropy has many applications as we will see in the following and constitutes the core of all things labeled ``information theory''. Let us thus look a bit closer at this quantity. 

\begin{lemma}
Let $X$ be some discrete random variable. Then its entropy $H(X)$ satisfies the two inequalities
	
\begin{equation*}
	0\leq H(X)\leq \log (n).
\end{equation*}

\end{lemma}

Note that the first inequality is a direct consequence of the properties of the information content and the second follows from Gibbs' inequality \cite{Cover1991}.

With regard to entropy, probability distributions having maximal entropy are often of interest in applications as they can be seen as the least restricted ones (i.e.\ having the least a priori assumptions), given the model parameters. The \emph{principle of maximum entropy} states that when choosing among a set of probability distributions with certain fixed properties, the preference should be given to distributions that have the maximal entropy among all considered distributions. This choice is justified as the one making the fewest assumptions on the shape of the distribution apart from the prescribed properties.  

For discrete probability distributions, the uniform distribution is the one with the highest entropy among all other distributions on the same base set. This can be well seen in the example in Figure~\ref{fig:binent}: The entropy of a Bernoulli distribution takes its maximum at $p=1/2$, the parameter value for which it corresponds to the uniform probability distribution on the two elements $0$ and $1$, each occurring with probability $1/2$.

For continuous, real-valued random variables with a given finite mean $\mu$ and variance $\sigma^2$, the normal distribution with mean $\mu$ and variance $\sigma^2$ has highest entropy. Demanding non-negativity and a non-vanishing probability on the positive real numbers (i.e., an infinite support) with positive given mean $\mu$ yields the exponential distribution with parameter $\lambda=1/\mu$ as a maximum-entropy distribution.     

\subsubsection*{Examples}

Before continuing, let us now compute some more entropies in order to get a feeling for this quantity. 

For a uniform probability distribution $P$ on $n$ events $\Omega=\{\omega_1,\dots,\omega_n\}$ each event has probability $P(\omega_i)=1/n$ and we obtain

\begin{equation*}
H(P)=-\sum_{i=1}^{n} \frac{1}{n} \log \frac{1}{n}= \log n,  
\end{equation*}

as the maximal entropy for all discrete probability distributions on the set $\Omega$.

Let us now compute the entropy of a Bernoulli random variable $X$, i.e.\ a binary random variable $X$ taking values $0$ and $1$ with probability $p$ and $1-p$, respectively. For the entropy of $X$ we get

\begin{equation*}
H(X)=-(p\log p + (1-p) \log (1-p)).
\end{equation*}

\begin{figure}
\centering
	\includegraphics{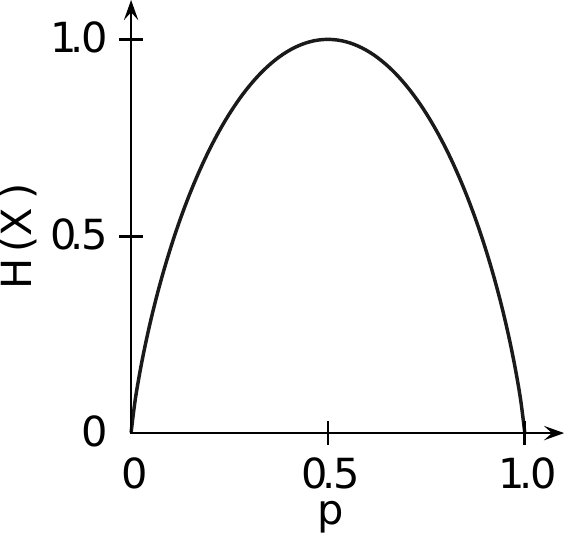}
\caption{Entropy $H(X)$ of a Bernoulli random variable $X$ as a function of success probability $p=P(X=1)$. The maximum is attained at $p=1/2$.}
\label{fig:binent}
\end{figure}

See Figure~\ref{fig:binent} for a plot of the entropy seen as a function of the success probability $p$. As expected, the maximum is attained at $p=1/2$, corresponding to the case of the uniform distribution.

Computing the differential entropy of a normal distribution $N(\mu,\sigma^2)$ with mean $\mu$ and variance $\sigma^2$ yields

\begin{equation*}
H(N(\mu,\sigma^2))=\frac{1}{2} \log (2\pi e \sigma^2),
\end{equation*}

and we see that the entropy does not depend on the mean value of the distribution but just its variance. This is not surprising, as the shape of the probability distribution is only changed by $\sigma^2$ and not $\mu$. 

For an example of how to compute the entropy of spike trains see Section~\ref{sec:neuralsystems}.

\subsubsection*{Joint Entropy}

Generalizing the notion of entropy to two or more variables we can define the so called joint entropy to quantify the expected uncertainty (or expected information) in a joint distribution of random variables.

\begin{definition}[joint entropy]
	Let $X$ and $Y$ be discrete random variables on some probability spaces. Then the \emph{joint entropy} of $X$ and $Y$ is given by
	
	\begin{equation}
	H(X,Y)=-E_{{X,Y}}[\log P(x,y)]=-\sum_{x,y}^{} P(x,y)\log P(x,y),
	\label{eq:defjointent}
	\end{equation}
	
	where $P_{X,Y}$ denotes the joint probability distribution of $X$ and $Y$ and the sum runs over all possible values $x$ and $y$ of $X$ and $Y$, respectively.	
\end{definition}

This definition allows a straightforward extension to the case of more than two random variables.

The conditional entropy $H(X|Y)$ of two random variables $X$ and $Y$ quantifies the expected uncertainty (respectively expected information) remaining in a random variable $X$ under the condition that a second variable $Y$ was observed or equivalently as the reduction of the expected uncertainty in $X$ upon the knowledge of $Y$. 

\begin{definition}[conditional entropy]
	Let $X$ and $Y$ be discrete random variables on some probability spaces. Then the \emph{conditional entropy} of $X$ given $Y$ is given by
	
	\begin{equation*}
	H(X|Y)=-E_{{X,Y}}[\log P(x|y)]=-\sum_{x,y}^{} P(x,y)\log P(x|y),
	\end{equation*}
	
	where $P_{X,Y}$ denotes the joint probability distribution of $X$ and $Y$.	
\end{definition}






\subsection{Mutual Information}
\label{sec:mutualinformation}

In this section we will introduce the notion of mutual information, an entropy-based measure for the information shared between two (or more) random variables. Mutual information can be thought of as a measure for the mutual dependence of random variables, i.e.\ as a measure for how far they are from being independent.

We will give two different approaches to this concept in the following: a direct one based on the point-wise mutual information $\operatorname{i}$ and one using the idea of conditional entropy. Note that in essence, these are just different approaches to defining the same object. We give the two approaches in the following, hoping that they help in understanding the concept better. In Section~\ref{sec:kullbackleibler} we will see yet another characterization in terms of the Kullback-Leibler divergence.


\subsubsection{Point-wise Mutual Information}

In terms of information content, the case of considering two events that are independent is straightforward: One of the axioms tells us that the information content of the two events occurring together is the sum of the information contents of the single events. But what about the case where the events non-independent? In this case we certainly have to consider the conditional probabilities of the two events occurring: If one event often occurs given that the other one occurs (think of the two events ``It is snowing.'' and ``It is winter''), the information overlap is higher than when the occurrence of one given the other is rare (think of ``It is snowing'' and ``It is summer.'').    

Using the notion of information from Section~\ref{ssec:information}, let us express this in a mathematical way by defining the \emph{mutual information} (i.e.\ shared information content) of two events. We call this the \emph{point-wise mutual information} or \emph{pmi}.

\begin{definition}[point-wise mutual information]
Let $x$ and $y$ be two events of a probability space $(\Omega, \Sigma, P)$. Then their \emph{point-wise mutual information} (pmi) is given as
\begin{equation}
\begin{aligned} 
\operatorname{i}(x;y) :&=  -\log\frac{P(x,y)}{P(x)P(y)}\\
&= -\log\frac{P(x|y)}{P(x)}\\
&= -\log\frac{P(y|x)}{P(y)}.
\label{eq:pmi}
\end{aligned} 
\end{equation}
\end{definition}

Note that we used joint probability distribution of $x$ and $y$ is for the definition of $\operatorname{i}(x;y)$ to avoid the ambiguities introduced by the conditional distributions. Yet, the latter are probably the easier way to gain a first understanding of this quantity.  

Let us note that this measure of shared information is symmetric ($\operatorname{i}(x;y)=\operatorname{i}(y;x)$) and that it can take any real value, particularly also negative values. Such negative values of point-wise mutual inforamtion are commonly referred to as \emph{misinformation} \cite{Lizier2008}. Point-wise mutual information is zero if the two events $x$ and $y$ are independent and it is bounded above by the information content of $x$ and $y$. More generally, the following inequality holds:  

\begin{equation*}
-\infty \leq \operatorname{i}(x;y) \leq \min\{ \underbrace{-\log P(x)}_{=h(x)}, \underbrace{-\log P(y)}_{=h(y)}\}. 
\end{equation*}

Defining the information content of the co-occurrence of $x$ and $y$ as $$\operatorname{i}(x,y):=-\log P(x,y),$$ another way of writing the point-wise mutual information is

\begin{equation}
\begin{aligned} 
\operatorname{i}(x;y) &= \operatorname{i}(x) + \operatorname{i}(y) - \operatorname{i}(x,y), \\ 
&= \operatorname{i}(x) - \operatorname{i}(x|y), \\
&= \operatorname{i}(y) - \operatorname{i}(y|x),
\end{aligned}
\label{eq:pmiconditional}
\end{equation}

where the first identity above is readily obtained from Equation~\ref{eq:pmi} by just expanding the logarithmic term and in the second and third line the formula for the conditional probability was used. 

Before considering mutual information of random variables as a straightforward generalization of the above, let us look at an example.

Say we have two probability spaces $(\Omega_a,\Sigma_a,P_a)$ and $(\Omega_b,\Sigma_b,P_b)$, with $\Omega_a=\{a_1,a_2\}$ and $\Omega_b=\{b_1,b_2\}$. We want to compute the point-wise mutual information of two events $\omega_a\in\Omega_a$ and $\omega_b\in\Omega_b$ subject to the joint probability distributions of $\omega_a$ and $\omega_b$ as given in Table~\ref{tab:pmitabomega}. Note that the joint probability distribution can also be written as matrix

\begin{equation*}
P(\omega_a,\omega_b)=\left(
\begin{tabular}{rr}
0.2&0.5\\
0.25&0.05
\end{tabular}
\right),
\end{equation*}

if we label rows by possible outcomes of $\omega_a$ and columns by possible outcomes of $\omega_b$. The marginal distributions $P(\omega_a)$ and $P(\omega_b)$ are now obtained as row, respectively column sums as $P(\omega_a=a_1)=0.7$, $P(\omega_a=a_2)=0.3$, $P(\omega_b=b_1)=0.45$, $P(\omega_b=b_2)=0.55$.   

We can now calculate the point-wise mutual information of for example

\begin{equation*}
\operatorname{i}(a_2;b_2)=-\log \frac{0.05}{0.3\cdot 0.55}\approx 1.7\ \mathrm{bits},
\end{equation*}

and

\begin{equation*}
\operatorname{i}(a_1;b_1)=-\log\frac{0.2}{0.7\cdot 0.45}\approx  -0.65\ \mathrm{bits}.
\end{equation*}

Note again that in contrast to mutual information (that we will discuss in the next section), point-wise mutual information can take negative values called, see\cite{Lizier2008} .


\begin{table}
\centering
\begin{tabular}{r|r|r}
$\omega_a$&$\omega_a$&P(x,y)\\\hline
$a_1$&$b_1$&0.2\\\hline
$a_1$&$b_2$&0.5\\\hline
$a_2$&$b_1$&0.25\\\hline
$a_2$&$b_2$&0.05
\end{tabular}
\caption{Table of joint probabilities $P(\omega_a,\omega_b)$ of two events $\omega_a$ and $\omega_b$.}
\label{tab:pmitabomega}
\end{table}











\subsubsection{Mutual Information as Expected Point-wise Mutual Information}

Using point-wise mutual information, the definition of mutual information of two random variables is straightforward: Mutual information of two random variables is the expected value of the point-wise mutual information of all realizations.

\begin{definition}[mutual information]
Let $X$ and $Y$ be two discrete random variables. Then the mutual information $I(X;Y)$ is given as the expected point-wise mutual information, \begin{equation}
\begin{aligned}
    I(X;Y):&= E_{{X,Y}}[\operatorname{i}(x;y)]\\ 
    &= \sum_{y} \sum_{x} P(x,y) \operatorname{i}(x;y)\\
    &= -\sum_{y} \sum_{x} P(x,y) \log{ \left(\frac{P(x,y)}{P(x)\,P(y)} \right) },
\end{aligned}
\end{equation}
where the sums are taken over all possible values $x$ of $X$ and $y$ of $Y$.
\end{definition}

Remember again that the joint probability $P(x,y)$ is just a two-dimensional matrix where the rows are indexed by $X$-values and the columns by $Y$-values and that each row (column) tells us how likely each possible value of $Y$ ($X$) is, given the value $x$ of $X$ ($y$ of $Y$) determined by the row (column) index. The rows (columns) sum to the marginal probability distributions $P(x)$ ($P(y)$), that can be written as vectors.  

If $X$ and $Y$ are continuous random variables we just replace the sums by integrals and obtain what is known as \emph{differential mutual information}:
\begin{equation} 
    I(X;Y) := \int_{\mathbb{R}} \int_{\mathbb{R}} P(x,y) \log{ \left(\frac{P(x,y)}{P(x)\,P(y)} \right) } \; \mathrm{d}x \,\mathrm{d}y. 
\end{equation}

Here $P(x,y)$ denotes the joint probability distribution function of $X$ and $Y$, and $P(x)$ and $P(y)$ the marginal probability distribution functions of $X$ and $Y$, respectively.

As we can see, mutual information can be interpreted as the information (i.e.\ entropy) shared by the two variables, hence its name. Like point-wise mutual information it is a symmetric quantity $I(X;Y)=I(Y;X)$ and it is non-negative, $I(X;Y)\geq 0$. Note though that it is not a metric, as in the general case it does not satisfy the triangle inequality. Furthermore we have $I(X;X)=H(X)$ and this identity is the reason why entropy is sometimes is also called \emph{self-information}.



\begin{figure}
\centering
	\includegraphics{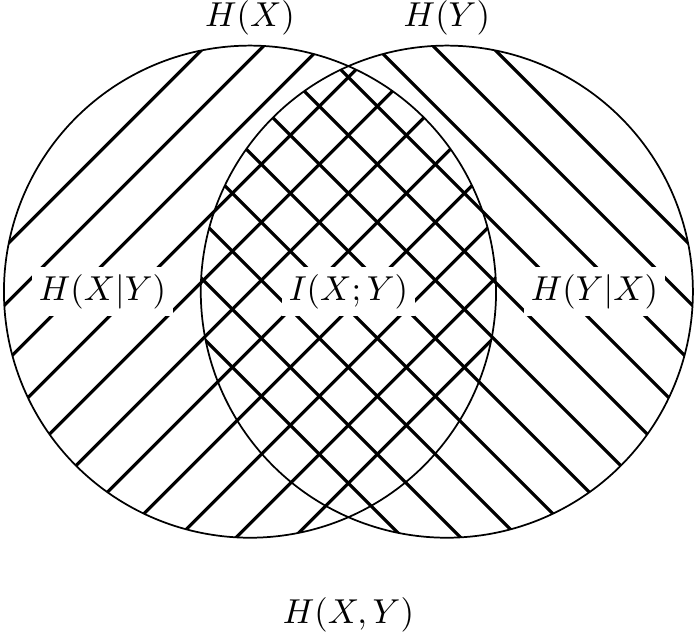}
\caption{Venn diagram showing the relation between the entropies $H(X)$ and $H(Y)$, the joint entropy $H(X,Y)$, the conditional entropies $H(X|Y)$ and $H(Y|X)$, and mutual information $I(X;Y)$.}
\label{fig:venndiagmi}
\end{figure}

Taking the expected value of Equation~\ref{eq:pmi} and using the notion of conditional entropy we can define mutual information between two random variables as follows.

\begin{equation}
\begin{aligned}
	I(X;Y):&=H(X)+H(Y)-H(X,Y),\\
	&=H(X)-H(X|Y),\\
	&=H(Y)-H(Y|X),
\end{aligned}
\label{eq:defmicondent}
\end{equation}

where in the last two steps the identity $H(X,Y)=H(X)+H(Y|X)=H(Y)+H(X|Y)$ was used. Note that Equation~\ref{eq:defmicondent} is the generalization of Equation~\ref{eq:pmiconditional} to the case of random variables. See Figure~\ref{fig:venndiagmi} for an illustration of how the relation between the different entropies and mutual information.

A possible interpretation of mutual information of two random variables $X$ and $Y$ is to consider it as a measure for the shared entropy between the two variables.

\subsubsection{Mutual Information and Channel Capacities}

We will look at channels in Shannon's sense of communication in the following and relate mutual information to channel capacity. But rather than looking at the subject in its full generality, we restrict ourselves to discrete, memoryless channels. The interested reader is pointed to \cite{Cover1991} for a more thorough treatment of the subject. 

Let us take as usual $X$ and $Y$ for the signal transmitted by some sender and received by some receiver, respectively. In terms of information transmission we can interpret mutual information $I(X;Y)$ as the average amount of information the received signal constrains about the transmitted signal, where the averaging is done over the probability distribution of the source signal $P_X$. This makes mutual information a function of $P_X$ and $P_{Y|X}$ and as we know, it is a symmetric quantity.

Shannon defines the \emph{capacity} $C$ of some channel as the maximum amount of information that a signal $Y$ received by the receiver can contain about the signal $X$ transmitted through the channel by the source.

In terms of mutual information $I(X;Y)$ we can define the channel capacity as the maximum mutual information $I(X;Y)$ among all realizations of the signal $X$. Channel capacity is thus not dependent on the distribution of $P_X$ of $X$ but rather a property of the channel itself, i.e.\ a property of the conditional distribution $P_{Y|X}$ and as such asymmetric and causal \cite{Wiener1956, Pearl2000}.  

Note that channel capacity is bound from below by $0$ and from above by the entropy $H(X)$ of $X$, with the maximal capacity being attained by a noise-free channel. In the presence of noise the capacity is lower.

We will have a look at channels again when dealing with applications of the theory in Section~\ref{sec:neuralsystems}.

\subsubsection{Normalized Measures of Mutual Information}

In many applications one is often interested in making values of mutual information comparable by employing a suitable normalization. Consequently, there exists a variety of proposed normalized measures of mutual information, most based on the simple idea of normalizing by one of the entropies that appear in the upper bounds of the mutual information. Using the entropy of one variable as a normalization factor, there a two possible choices and both were proposed: The so called \emph{coefficient of constraint} $C(X|Y)$ \cite{Coombs1970} 

\begin{equation*}
     C(X|Y):=\frac{I(X;Y)}{H(Y)}  
\end{equation*}

and the \emph{uncertainty coefficient} $U(X|Y)$ \cite{Theil1992}

\begin{equation*}
     U(X|Y):=\frac{I(X;Y)}{H(X)}.  
\end{equation*}

These two quantities are obviously non-symmetric but can easily be symmetrized for example by setting

\begin{equation*}
U(I,J) := \frac{H(I)U(I|J)+H(J)U(J|I)}{H(I)+H(J)}. 
\end{equation*}

Another symmetric normalized measure for mutual information, usually referred to as \emph{redundancy measure}, is obtained when normalizing using the sum of the entropy of the variables

\begin{equation*}
     R= \frac{I(X;Y)}{H(X)+H(Y)}.
\end{equation*}

Note that $R$ takes its minimum of $0$ when the two variables are independent and its maximum when one variable is completely redundant knowing the other. 

Note that the list of normalized variants of mutual information given here is far from complete. But as said earlier, the principle behind most normalizations is to use one or a combination of the entropies of the involved random variables as a normalizing factor.  



\subsubsection{Multivariate Case}

What if we want to calculate the mutual information between not only between two random variables but rather three or more? A natural generalization of mutual information to this so called \emph{multivariate} case is given by the following definition using conditional entropies and is also called \emph{multi-information} or \emph{integration} \cite{Tononi1994}. 

The mutual information of three random variables $X_1,\dots,X_3$ is given by

\begin{equation*}
I(X_1;X_2;X_3) := I(X_1;X_2) - I(X_1;X_2|X_3),
\end{equation*}

where the last term is defined as 

\begin{equation*}
I(X_1;X_2|X_3):= E_{X_3}\left[I(X_1;X_2)|X_3\right],
\end{equation*}

thee latter being called the \emph{conditional mutual information} of $X_1$ and $X_2$ given $X_3$.  The conditional mutual information $I(X_1;X_2|X_3)$ can also be interpreted as the average common information shared by $X_1$ and $X_2$ that is not already contained in $X_3$.

Inductively, the generalization to the case of $n$ random variables  $X_1,\dots,X_n$ is straightforward:   

\begin{equation*}
I(X_1;\dots;X_n) := I(X_1;\dots;X_{n-1}) - I(X_1;\dots;X_{n-1}|X_n),
\end{equation*}

where the last term is again the conditional mutual information 

\begin{equation*}
I(X_1;\ldots;X_{n-1}|X_{n}) := E_{X_{n}}\left[I(X_1;\ldots;X_{n-1})|X_{n}\right]. 
\end{equation*}

Beware that while the interpretations of mutual information directly generalize from the bi-variate case $I(X;Y)$ to the multivariate case $I(X_1;\dots;X_n)$ there is an important difference between the bivariate and the multivariate measure. Whereas mutual information $I(X;Y)$ is a non-negative quantity, multivariate mutual information (MMI for short) behaves a bit differently than the usual mutual information in the aspect that it can also take negative values which makes this information-theoretic quantity sometimes difficult to interpret.

Let us first look at an example of three variables with positive MMI. To make things a bit more hands on, let us look at three binary random variables, one telling us whether it is cloudy, the other whether it is raining and the third one whether it is sunny. We want to compute $I({\rm rain};{\rm no\ sun};{\rm cloud})$. In our model, clouds can cause rain and can block the sun and so we have

\begin{equation*}
I({\rm rain};{\rm no\ sun}|{\rm cloud})\leq I({\rm rain};{\rm no\ sun}),
\end{equation*}

as it is more likely that it is raining and there is no sun visible when it is cloudy than when there are no clouds visible. This results in positive MMI for $I({\rm rain};{\rm no\ sun};{\rm cloud})$, a typical situation for a common-cause structure in the variables: here, the fact that the sun is not shining can partly be due to the fact that it is raining and partly due to the fact that there are clouds visible.    

In a sense the inverse is the situation where we have two causes with a common effect: This situation can lead to negative values for the MMI, see \cite{MacKay2003}. In this situation, observing a common effect induces a dependency between the causes that did not exist before. This fact is called ``explaining away'' in the context of Bayesian networks, see \cite{Pearl1988}. Pearl \cite{Pearl1988} also gives a car-related example where the three (binary) variables are ``engine fails to start'' ($X$), ``battery dead'' ($Y$) and ``fuel pump broken'' ($Z$). Clearly, both $Y$ and $Z$ can cause $X$ and are uncorrelated if we have no knowledge of the value of $X$. But fixing the common effect $X$, namely observing that the engine did not start, induces a dependency between $Y$ and $Z$ that can lead to negative values of the MMI.

Another problem with the $n$-variate case to keep in mind is the combinatorial explosion of the degrees of freedom regarding their interactions. As a priori every non-empty subset of the variables could interact in an information-theoretic sense, this yields $2^{n-1}$ degrees of freedom.

\subsection{A Distance Measure for Probability Distributions: the Kullback-Leibler Divergence}
\label{sec:kullbackleibler}

The \emph{Kullback-Leibler divergence} \cite{Kullback1951} (or \emph{KL-divergence} for short)  is a kind of ``distance measure'' on the space of probability distributions: Given two probability distributions on the same base space $\Omega$ interpreted as two points in the space of all probability distributions over the base set $\Omega$, it tells us how far they are ``apart''. 

We again use the usual expectation-value construction as used for the entropy before.

\begin{definition}[Kullback-Leibler divergence]
	Let $P$ and $Q$ be two discrete probability distributions over the same base space $\Omega$. Then the \emph{Kullback-Leibler divergence} of $P$ and $Q$ is given by
	
	\begin{equation}
		D_{\mathrm{KL}}(P\|Q):=\sum_{\omega\in\Omega} P(\omega) \log \frac{P(\omega)}{Q(\omega)}.
		\label{eq:kldefinition}
	\end{equation}

\end{definition}

The Kullback-Leibler divergence is non-negative $D_{\mathrm{KL}}(P\|Q) \geq 0$ (and it is zero if $P$ equals $Q$ almost everywhere), but it is not a metric in the mathematical sense as in general it is non-symmetric $D_{\mathrm{KL}}(P\|Q)\neq D_{\mathrm{KL}}(Q\|P)$ and it does not fulfill the triangle inequality. Note that in their original work, Kullback and Leibler \cite{Kullback1951} defined the divergence via the sum 

\begin{equation*}
    D_{\mathrm{KL}}(P\|Q) + D_{\mathrm{KL}}(Q\|P), 
\end{equation*}

making it a symmetric measure. $D_{\mathrm{KL}}(P\|Q)$ is additive for independent distributions, namely  

\begin{equation*}
D_{\mathrm{KL}}(P \| Q) = D_{\mathrm{KL}}(P_1 \| Q_1) + D_{\mathrm{KL}}(P_2 \| Q_2), 
\end{equation*}

where the two pairs $P_1, P_2$ and $Q_1, Q_2$ are independent probability distributions with the joint distributions $P=P_1P_2$ and $Q=Q_1Q_2$, respectively.

Note that the expression in Equation~\ref{eq:kldefinition} is nothing else than the expected value $E_P[\log P - \log Q]$ with the expectation value taken with respect to $P$, which in term can be interpreted as ``expected distance of $P$ and $Q$'', measured in terms of the information content. Another interpretation can be given in the language of codes: $D_{\mathrm{KL}}(P\|Q)$ is the average number of extra bits needed to code samples from $P$ using a code book based on $Q$.

Analogous to previous examples, the KL-divergence can also be defined for continuous random variables in a straightforward way via

\begin{equation*}
    D_{\mathrm{KL}}(P\|Q) = \int_{\mathbb{R}} p(x)\log\left(\frac{p(x)}{q(x)}\right) \; \mathrm{d}x , 
\end{equation*}

where $p$ and $q$ denote the pdf of two continuous probability distributions $P$ and $Q$.

Expanding the logarithm in Equation~\ref{eq:kldefinition} we can write the Kullback-Leibler divergence between two probability distributions $P$ and $Q$ in terms of entropies as

\begin{equation*}
D_{\mathrm{KL}}(P\|Q) = - \operatorname{E}_P(\log q(x)) + \operatorname{E}_P(\log p(x)) = H^{\mathrm{cross}}(P,Q) - H(P),
\end{equation*}

where $p$ and $q$ denote the pdf or pmf of the distributions $P$ and $Q$ and $H(P,Q)^{\mathrm{cross}}$ is the so-called \emph{cross-entropy} of $P$ and $Q$ given by

\begin{equation*}
\mathrm{H}^{\mathrm{cross}}(P, Q) := -E_P(\log Q).
\end{equation*}

\begin{figure}
\centering
\subfigure[]{%
\label{fig:klnorm1}%
\includegraphics{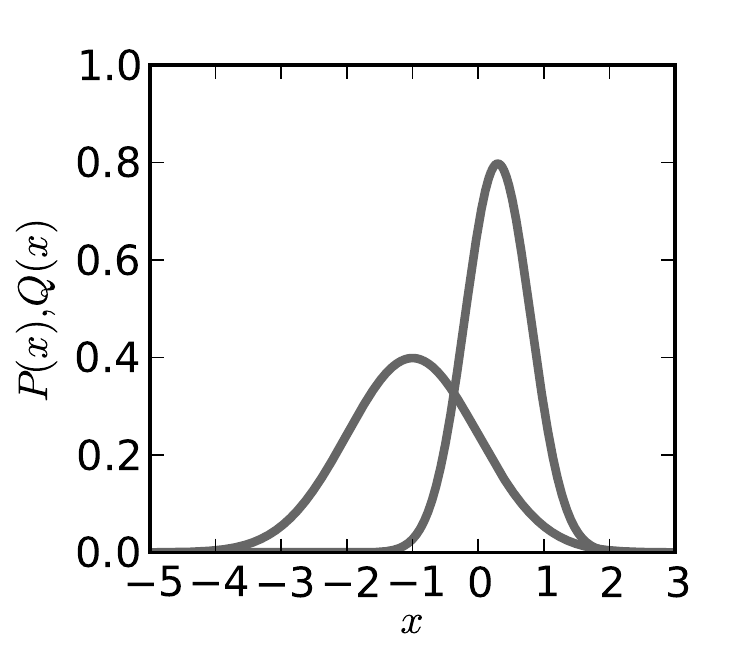}
}
\subfigure[]{%
\label{fig:klnorm2}%
\includegraphics{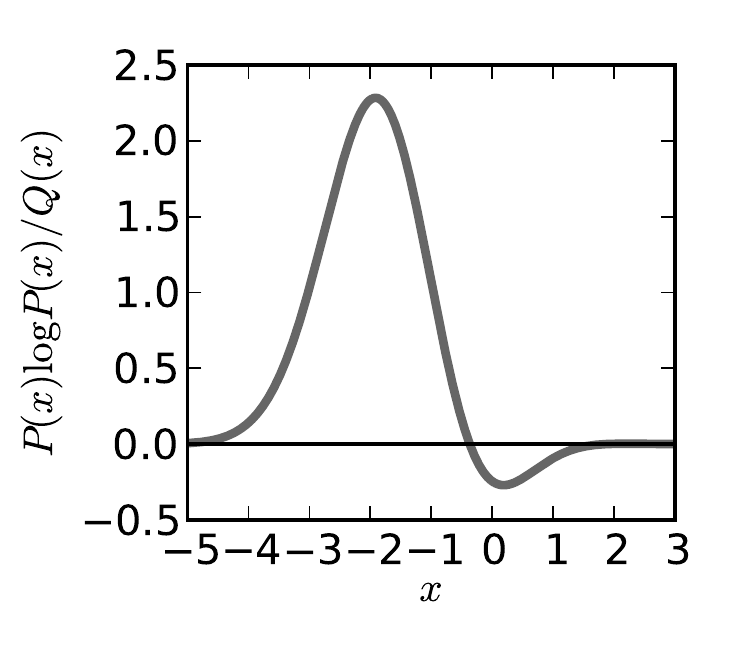}
}
\caption{The probability densities of two Gaussian probability distributions \subref{fig:klnorm1} and the quantity $P(x)\log P(x)/Q(x)$ that yields the KL-Divergence when integrated \subref{fig:klnorm2}.}
\label{fig:klnorm}
\end{figure}

This relation lets us easily compute a closed form of the KL-Divergence for many common families of probability distributions. Let us for example look at the value of the KL-Divergence between two normal distributions $P\sim N(\mu_1,\sigma_1^2)$ and $Q\sim N(\mu_2,\sigma_2^2)$, see Figure~\ref{fig:klnorm}. This can be calculated as 

\begin{equation*}
D_\mathrm{KL}( P \| Q ) = \frac{(\mu_1 - \mu_2)^2}{2\sigma_2^2} + \frac12\left( \frac{\sigma_1^2}{\sigma_2^2}  - \log\frac{\sigma_1^2}{\sigma_2^2} - 1 \right). 
\end{equation*}

Another example: The KL-divergence between two exponential distributions $P\sim \operatorname{Exp}(\lambda_1)$ and $Q\sim \operatorname{Exp}(\lambda_2)$ is

\begin{equation*}
D_\mathrm{KL}( P \| Q ) = \log(\lambda_1) - \log(\lambda_2) + \frac{\lambda_2}{\lambda_1} - 1. 
\end{equation*}

Using the Kullback-Leibler divergence we can give yet another characterization of mutual information: It is a measure of how far two measured variables are from being independent, this time in terms of the Kullback-Leibler divergence.

\begin{equation}
	\begin{aligned}
		I(X;Y)&=H(X)-H(X|Y)\\
		&=-\underbrace{\sum_{x} P(x)\log(P(x))}_{=-\sum_{x,y} P(x,y)\log(P(x))}+ \sum_{x,y} P(x,y)\log(P(x|y))\\
		&=\sum_{x,y} P(x,y)\log(\frac{P(x|y)}{P(x)})\\
		&=\sum_{x,y} P(x,y)\log(\frac{P(x,y)}{P(x)P(y)})\\
		&=D_{\mathrm{KL}}(P(x,y)\|P(x)P(y))\\
	\end{aligned}
\end{equation}

Thus, mutual information of two random variables can be seen as the KL-Divergence of their underlying joint probability distribution from the products of their marginal probability distributions, i.e.\ as a measure for how far the two variables are from being independent.

\subsection{Transfer Entropy: Conditional Mutual Information}
\label{sec:transferentropy}

In the past, mutual information was often used as a measure of information transfer between units (modeled as random variables) in some system. This approach faces the problem that mutual information is a symmetric measure and does not have an inherent directionality. In some applications this symmetry is not desired though, namely whenever we want to explicitly obtain information about the ``direction of flow'' of information, for example to measure causality in an information-theoretic setting, see Section~\ref{sec:causality}. 

In order to make mutual information a directed measure, a variant called \emph{time-lagged mutual information} was proposed, calculating mutual information for two variables including a previous state of the source variable and a next state of the destination variable (where discrete time is assumed).

Yet, as Schreiber \cite{Schreiber00MeasInfTrans} points out, while time-lagged mutual information provides a directed measure of information transfer, it does not allow for a time-dynamic aspect as it measures the statically shared information between the two elements. With a suitable conditioning on the past of the variables, the introduction of a time-dynamic aspect is possible though. The resulting quantity is commonly referred to as \emph{transfer entropy} \cite{Schreiber00MeasInfTrans}. Its common definition is the following.

\begin{definition}[transfer entropy]
Let $X$ and $Y$ be discrete random variables given on a discrete time scale and let $k,l\geq 1$ be two natural numbers. Then the \emph{transfer entropy from $Y$ to $X$ with $k$ memory steps in $X$ and $l$ memory steps in $Y$} is defined as

\begin{equation*}
\operatorname{TE}_{Y\rightarrow X}: \sum_{x_{n+1},x_n^k, y_n^l} P(x_{n+1},x_n^k,y_n^l)\log \frac{P(x_{n+1}|x_n^k,y_n^l)}{P(x_{n+1}|x_n^k)} ,	
\end{equation*}

where we denoted by $x_n, y_n$ the value of $X$ and $Y$ at time $n$ and by $x_n^k$ the past $k$ values of $X$, counted from time $n$ on: $x_n^k:=(x_n,x_{n-1},\dots,x_{n-k+1})$, and analogously $y_n^l:=(y_n,y_{n-1},\dots,y_{n-l+1})$.  
\end{definition}

Although this definition might look complicated at first, the idea behind it is quite simple. It is merely the Kullback-Leibler divergence between the two conditional probability distributions $P(x_{n+1}|x_n^k)$ and $P(x_{n+1}|x_n^k,y_n^l)$,

\begin{equation*}
\operatorname{TE}_{Y\rightarrow X}=D_{\mathrm{KL}}(P(x_{n+1}|x_n^k)\| P(x_{n+1}|x_n^k,y_n^l)), 
\end{equation*}

i.e.\ a measure of how far the two distributions are from fulfilling the generalized Markov property (see Section~\ref{sec:stochasticproc})

\begin{equation}
P(x_{n+1}|x_n^k)=P(x_{n+1}|x_n^k,y_n^l).
\label{eq:tegenmarkov}
\end{equation}

Note that for small values of transfer entropy, we can say that $Y$ has little influence on $X$ at time $t$, whereas we can say that information is transferred from $Y$ to $X$ at time $t$ when the value is large. Yet, keep in mind that transfer entropy is just a measure of statistical correlation, see Section~\ref{sec:causality}.

Another interpretation of transfer entropy is seeing it as a conditional mutual information $I(Y^{(l)};X'|X^{(k)}$, measuring the average information the source $Y$ constrains about the next state $X'$ of the destination $X$ that was not contained in the destination's past $X^{(k)}$ (see \cite{Lizier2010}) or alternatively as the average information provided by the source about the state transition in the destination, see \cite{Lizier2010, Kaiser2002}.


As so often before, the concept can be generalized to the continuous case \cite{Kaiser2002}, although the continuous setting introduces some subtleties that have to be addressed.

\begin{figure}
\centering
\includegraphics{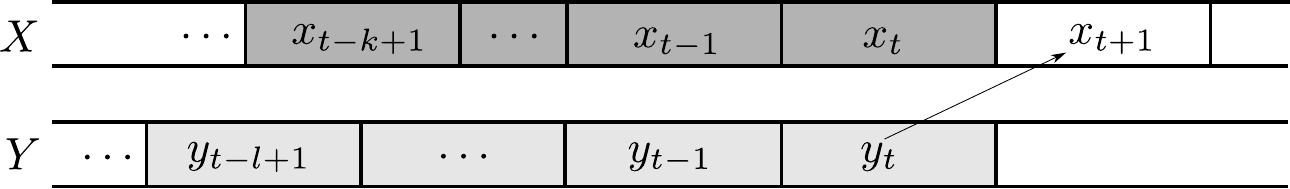}
\caption{Computing transfer entropy $\operatorname{TE}_{Y\rightarrow X}$ from source $Y$ to target $X$ at time $t$ as a measure of the average information present in $y_t$ about the future state $x_{t+1}$. The memory vectors $x_n^k$ and $y_n^k$ are shown in gray.}
\label{fig:te}
\end{figure}

Concerning the memory-parameters $k$ and $l$ of the source and the destination, although arbitrary choices are possible, the values chosen fundamentally influence the nature of the questions asked. In order to get correct measures for systems being far from Markovian (i.e.\ systems which states are not influenced by more than a certain fixed number of preceding system states), high values of $k$ have to be used, and for non-Markovian systems the case $k\rightarrow \infty$ has to be considered. On the other hand, commonly just one previous state of the source variable is considered, setting $l=1$ \cite{Lizier2010}, this being also due to the growing data intensity in $k$ and $l$ and the usually high computational cost of the method.    

Note that akin to the case of mutual information, there exist point-wise versions of transfer entropy (also called \emph{local transfer entropy}), as well as extensions to the multivariate case, see \cite{Lizier2010}. 

\section{Estimation of Information-theoretic Quantities}
\label{sec:estimation}

As we have seen in the preceding sections, one needs to know the full sample spaces and probability distributions of the random variables involved in order to precisely calculate information-theoretic quantities such as the entropy, mutual information or transfer entropy. But obtaining this data is in most cases impossible in reality, as the spaces are usually high-dimensional and sparsely sampled, rendering the direct methods for the calculation of such quantities impossible to carry out. 

A way around this problem is to come up with estimation techniques that estimate entropies and derived quantities such as mutual information from the data. Over the last decades a large body of research was published concerning the estimation of entropies and related quantities, leading to a whole zoo of estimation techniques, each class having its own advantages and drawbacks. So rather than a full overview, we will give a sketch of some central ideas here and give references to further literature. The reader is also pointed to the review articles \cite{Beirlant1997, Paninski2003}.

Before looking at estimation techniques for neural (and other) data let us first give a swift and painless review some important theoretical concepts regarding statistical estimation.

\subsection{A Bit of Theory Regarding Estimations}

From a statistical point of view, the process of estimation in its most general form can be regarded in the following setting: We have some data (say measurements or data obtained via simulations) that is believed to be generated by some stochastic process with an underlying non-autonomous, i.e. time-dependent, or autonomous probability distribution. We then want to estimate either the value of some function defined on that probability distribution (for example the entropy) or the shape of this probability distribution as a whole (from which we can then obtain an estimate of a derived quantity). This process is called \emph{estimation} and a function mapping the data to an estimated quantities estimator. In this section will will first look at estimators and their desired properties and then look at what is called maximum likelihood estimation, the most commonly used method for the estimation of parameters in the field of statistics.

\subsubsection{Estimators}

Let $x=(x_1,\dots,x_n)$ be a set of realizations of the random variable $X$ that is believed to have a probability distribution that comes from a family of probability distributions $P_{\theta}$ parametrized by a parameter $\theta$ and assume that the underlying probability distribution of $X$ is $P_{\theta_{\mathrm{true}}}$.

Let $T: x\mapsto \hat{\theta}_{\mathrm{true}}$ be an estimator for the parameter $\theta$ with the true value $\theta_{\mathrm{true}}$. For the value of the estimated parameter we usually write $\hat{\theta}_{\mathrm{true}}:=T(x)$. The \emph{bias} of $T(x)$ is the expected difference between $\hat{\theta}_{\mathrm{true}}$ and $\theta_{\mathrm{true}}$: $$\operatorname{bias}(T):=E_X[\hat{\theta}_{\mathrm{true}}-\theta_{\mathrm{true}}],$$ 
and an estimator with vanishing bias is called \emph{unbiased}. 

We usually also want the estimator to be \emph{consistent}, i.e.\ we want the estimated value $\hat{\theta}_{\mathrm{true}}$ to converge to the value of the true parameter $\theta_{\mathrm{true}}$ in probability as the sample $x$ increases in size, i.e.\ as $n\rightarrow \infty$:
$$\lim_{n \to \infty}P(|T(X) -\theta_{\mathrm{true}}|>\varepsilon)=0.$$

Another important property of an estimator is its variance $\operatorname{var}(T)$ and an unbiased estimator having the minimal variance among all unbiased estimators of the same parameter is called \emph{efficient}.   

Yet another measure often used when assessing the quality of an estimator $T$ is its mean squared error

$$\operatorname{MSE}(T) = \left( \operatorname{bias}(T) \right)^{2}+\operatorname{var}(T)$$

and as we can see, any unbiased estimator with minimal variance minimizes the mean squared error.

Without further going into detail here, it is noted that there exists a  theoretical lower bound to the minimal variance obtainable by an unbiased estimator, the \emph{Cram\'{e}r-Rao bound}. The Cram\'{e}r-Rao bound sets the variance of the estimator in relation with the so called \emph{Fisher information} (that can be set into relation with mutual information, see \cite{Brunel1998, Yarrow2012}). The interested reader is pointed to \cite{Lehmann1998, Amari00MethInfGeo}.

\subsubsection{Estimating Parameters: The Maximum Likelihood Estimator}

\emph{Maximum likelihood estimation} is the most-widely used estimation technique in statistics and, as we will see in the next few paragraphs, a straightforward procedure that in essence tells us what the most likely parameter value in an assumed family of probability distributions is, given a set of realizations of a random variable that is believed to have a underlying probability distribution from the family considered. 

In statistical applications one often faces the following situation: We have a finite set of realizations $\{x_i\}_i$ of a random variable $X$. We assume $X$ to have a probability distribution $f(x,\theta_{\mathrm{true}})$ in a certain parametrized class of probability distributions $\{f(x,\theta)\}_{\theta}$, where the true parameter $\theta_{\mathrm{true}}$ is unknown. The goal is to get an estimate $\hat{\theta}_{\mathrm{true}}$ of $\theta_{\mathrm{true}}$ using the realizations $\{x_i\}_i$, i.e.\ to do statistical inference of the parameter $\theta$. Let us consider the so called \emph{likelihood function}

\begin{equation*}
L(\theta|x)=P_{\theta}(X=x)=f(x|\theta)
\end{equation*}

as a function of $\theta$. It is a measure of how likely it is that the parameter of the probability distribution has the value $\theta$, given the observed realization $x$ of $X$. In maximum likelihood estimation, we look for the parameter that maximizes the likelihood function. This is $\hat{\theta}_{\mathrm{true}}$:

\begin{equation*}
\hat{\theta}_{\mathrm{true}}=\operatorname{argmax}_{\theta} L(\theta|x).
\end{equation*}

Choosing a value of $\theta=\hat{\theta}_{\mathrm{true}}$ minimizes the KL-divergence between $P_{\theta}$ and $P_{\theta_{\mathrm{true}}}$ for all possible values of $\theta$. The value $\hat{\theta}_{\mathrm{true}}$, often written as $\hat{\theta}_{\mathrm{MLE}}$ is called the \emph{maximum likelihood estimate} (MLE for short) of $\theta_{\mathrm{true}}$.

In this setting, one often not uses the likelihood function directly, but works with the $\log$ of the likelihood function (this is referred to as log-likelihood). Why? The likelihood functions are often very complicated and situated in high dimensions, making it impossible to find a maximum of the function analytically. Thus, numerical methods (such as Newton's method and variants or the simplex method) have to be employed in order to find a solution. These numerical methods work best (and can be shown to converge to a unique solution) if the function they operate on is concave (bowl-shaped, where the closed end is on the top). The log-function has the property to make the likelihood function concave in many cases, that being the reason why one considers the log-likelihood function, rather than the likelihood function directly, see also \cite{Paninski2004a}.

\subsection{Regularization}

Having looked at some core theoretical concepts regarding the estimation of quantities depending on probability distributions let us now come back to dealing with real data.

As in real-world data, the involved probability distributions are often continuous and infinite-dimensional, the resulting estimation problem is very difficult (if not impossible) to solve in its original setting. As a remedy, the problem is often \emph{regularized}, i.e.\ mapped to a discrete, more easily solvable problem. This of course introduces errors and often makes a direct estimation of the information-theoretic quantities impossible, but even in that simplified model we can estimate lower bounds of the quantities that By using Shannon's \emph{information processing inequality} \cite{Cover1991}

\begin{equation*}
I(X;Y)\geq I(S(X);T(Y)),
\end{equation*}

where $X$ and $Y$ are (discrete) random variables and $S$ and $T$ are measurable maps.

By choosing the mappings $S$ and $T$ as our regularization mappings (you might also regard them as parameters) we can change the coarseness of the regularization. The regularization can be chosen arbitrarily coarse, i.e.\ choosing $S$ and $T$ as constant functions, but this of course comes with a price. For example in the latter case of constant $S$ and $T$ the mutual information $I(S(X);S(Y))$ would be equal to $0$, clearly not a very useful estimate. This means that a trade-off between complexity reduction and the quality of the estimation has to be made. In general, there exists no all-purpose recipe for this, each problem requiring an appropriate regularization.   

As this discretization technique has become the standard method in many fields, we will solely consider the regularized, discrete case in the following and point the reader to the review article \cite{Beirlant1997} concerning the continuous case.

In the neurosciences, such a regularization technique was also proposed and is known as the ``direct method'' \cite{Strong1998, Buracas1998}. Here, spike trains of recorded neurons are discretized into time bins of a given fixed width and the neuronal spiking activity is interpreted as a series of symbols from an alphabet defined via the observed spiking pattern in the time bins. 

\subsection{Non-parametric Estimation Techniques}

Commonly, two different classes of estimation techniques regarding the shape of probability distributions are distinguished. Parametric estimation techniques assume that the probability distribution is contained in some family of probability distributions having some prescribed shape (see Section~\ref{sec:paramprobdistr}). Here, one estimates the value of the parameter from the data observed, whereas non-parametric estimation techniques make no assumptions about the shape of the underlying distribution. We will solely look at non-parametric estimation techniques in the following as in many cases one tries to not assume prior information about the shape of the distribution. 



Histogram-based estimation is the most popular and most-widely used  estimation technique. As the name implies, this method uses a histogram obtained from the data to estimate the probability distribution of the underlying random generation mechanism. 


For the following, assume that we obtained a finite set of $N$ samples $x=\{x_i\}_i$ of some real random variable $X$ defined over some probability space $(\Omega,\Sigma,P)$. We then divide the domain of $X$ into $m\in \mathbb{N}$ equally sized bins $\{b_i\}_i$ and subsequently count the number of realizations $x_i$ in our data set contained in each each bin. Here, the number $m$ of bins can be freely chosen. It controls the coarseness of our discretization, where the limit $m\rightarrow \infty$ is the continuous case. This allows us to define relative frequencies of occurrences for $X$ with respect to each bin that we interpret as estimations $\hat{p}_i^m$ (note that we make the dependence on the number of bins $m$ explicit in the notation) of the probability of $X$ taking a value in bin $b_i$ which we denote by $p_i^m=P(X\in b_i)$. The law of large numbers then tells us that our estimated probability values converge to the real probabilities as $N\rightarrow \infty$. 

Note that although histogram-based estimations are usually called non-parametric as they do not assume a certain shape of the underlying probability distribution, they do have parameters, namely one parameter for each bin, the estimated probability value $\hat{p}_i^m$. These estimates  $\hat{p}_i^m$ can also be interpreted as maximum-likelihood estimates of $p_i$.

The following defines an estimator of the entropy based on the histogram. It is often called ``plug in'' estimator:

\begin{equation}
\hat{H}_{\mathrm{MLE}}(x):=-\sum_{i=1}^{m} \hat{p}_i^m \log p_i^m.
\label{eq:hxmle}
\end{equation}

\begin{figure}
\centering
	\includegraphics{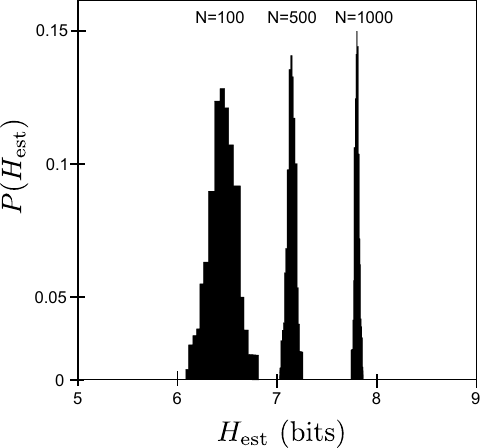}
\caption{Estimation bias for a non-bias corrected histogram-based maximum likelihood estimator $H_{\mathrm{est}}$ of the entropy of a given distribution with true entropy $H=8$ bits. Estimated values are shown for three different sample sizes $N$. Adapted from \cite{Paninski2003}, Figure~1.}
\label{fig:hest}
\end{figure}

The are some problems with this estimator $\hat{H}_{\mathrm{MLE}}(X)$, though. Its convergence to the true value $H(X)$ can be slow and it is negatively biased, i.e.\ its value is almost always below the true value $H(X)$, see \cite{Panzeri1995, Antos2001, Paninski2003, Panzeri2007}. This shift can be quite significant even for large $N$, see Figure~\ref{fig:hest} and \cite{Paninski2003}. More specifically, one can show that the expected value of the estimated entropy is always smaller than the true value of the entropy, 

\begin{equation*}
E_X[\hat{H}_{\mathrm{MLE}}(x)]\leq H(X),
\end{equation*}

where the expectation value is taken with respect to the true probability distribution $P$.

Bias generally is a problem for history-based estimation techniques \cite{Blyth1959, Schurmann2004, Panzeri2007} and although we can correct for the bias, this may not always be a feasible solution \cite{Antos2001}. None the less we will have a look at a bias-corrected version of the estimator given in Equation~\ref{eq:hxmle} below.   

As a remedy to the bias problem, Miller and Madow \cite{Miller1955} calculated the bias of the estimator of Equation~\ref{eq:hxmle} and came up with a bias-corrected version of the maximum likelihood estimator for the entropy, referred to as \emph{Miller-Madow} estimator:

\begin{equation*}
\hat{H}_{\mathrm{MM}}(x):=\hat{H}_{\mathrm{MLE}}(x) +\frac{\hat{m}-1}{2N},
\end{equation*}

where $\hat{m}$ is an estimate of the number of bins with non-zero probability. We will not go into the detail of the method here, the interested reader is referred to \cite{Miller1955}.

Another way of bias-correction $\hat{H}_{\mathrm{MLE}}(X)$ is the so called ``jack-knifed'' version of the maximum-likelihood estimator by Efron and Stein \cite{Efron1981}:

\begin{equation*}
\hat{H}_{\mathrm{JK}}(x):=N\cdot \hat{H}_{\mathrm{MLE}}(x) +\frac{N-1}{N}\sum_{j=1}^{N} \hat{H}_{\mathrm{MLE}}(x\backslash \{x_i\}),
\end{equation*}

Yet another bias-corrected variant of the MLE estimator based on polynomial approximation is presented in \cite{Paninski2003}, for which also bounds on the maximal estimation error were derived. 


In an effort to overcome the problems faced by histogram-based estimation, many new and more powerful estimation techniques have emerged over the last years, both for entropy and other information-theoretic quantities. As our focus here is to give an introduction to the field, we will not review all of those methods here but rather point the interested reader to the literature, where a variety of approaches are discussed. There exist methods based on the idea of adaptive partitioning of sample space \cite{Cellucci2005}, ones using entropy production rates and allowing for confidence intervals \cite{Shlens2007}, ones using Bayesian methods \cite{Nemenman2004, Shlens2007, Rad2011} and ones based on density estimation using nearest-neighbors \cite{Kraskov04EstMutInf}, along with many more. See \cite{Hlavackovaschindler2007} for an overview concerning several estimation techniques for entropy and mutual information. We note here that in contrast to estimations of entropy, estimators of mutual information are usually positively biased, i.e.\ tend to overestimate mutual information.











\section{Information-theoretic Analyses of Neural Systems}
\label{sec:neuralsystems}

Some time after its discovery by Shannon, neuroscientists started to recognize information theory as a valuable mathematical tool to assess information processing in neural systems. Using information theory, several questions regarding information processing and the neural code can be addressed in a quantitative way, among those

\begin{itemize}
	\item how much information single cells or populations carry about a stimulus and how this information is coded,
	\item what aspects of a stimulus are encoded in the neural system and
	\item how ``effective connectivity'' \cite{Friston1994} in neural systems can be defined via causal relationships between units in the system.
\end{itemize}

\begin{figure}
\centering
	\includegraphics{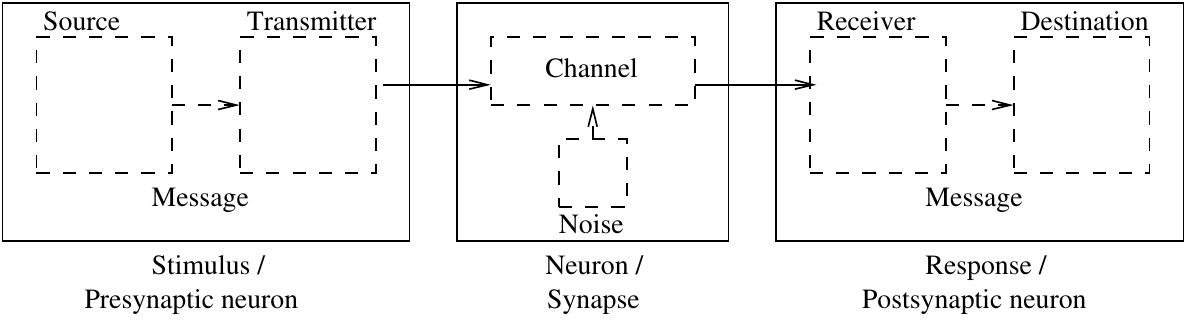}
\caption{An information-theoretic view on neural systems. Neurons can either act as channels in the information-theoretic sense, relaying information about some stimulus or as senders and receivers with channels being synapses.}
\label{fig:shannonchannel-neuron}
\end{figure}

See Figure~\ref{fig:shannonchannel-neuron} for an illustration of how Shannon's theory can be used in a neural setting.

Attneave \cite{Attneave1954} and Barlow \cite{Barlow1961} were the first to consider information processing in neural systems from an information-theoretic point of view. Subsequently, Eckhorn and Pöpel~\cite{Eckhorn1974, Eckhorn1975} applied information-theoretic methods to electrophysiologically recorded data of neurons in a cat. But being data-intensive in nature these methods faced some quite strong restrictions during that time, namely the limited amount of computing power (and computer memory) and the limited amount (and often low quality) of data obtainable via measurements at that time.

But over the last decades, available computing became more and more available and classical measurement techniques were improved, along with new ones emerging such as fMRI, MEG and calcium imaging. This made information theoretic analyses of neural systems more and more feasible and through the invention of recording techniques such as MEG and fMRI it is nowadays even possible to perform such analyses on a system-scale for the human brain in vivo. Yet, even with the newly available recording techniques today there are some conceptual difficulties with information-theoretic analyses as it is often a challenge to obtain enough data in order to get good estimates of information-theoretic quantities. Special attention has to be paid to using the data efficiently and the validity of such analyses has to be assessed to their statistical significance.

In the following we will discuss some conceptual questions relevant when regarding information theoretic analyses of neural systems. More detailed reviews can be found in \cite{Borst1999, Victor2006, Rolls2011, Fairhall2012}. 

\subsection{The Question of Coding}

Marr described ``three levels at which any machine carrying out an information-processing task must be understood'' \cite{Marr82Vision}[Chapter~1.2]. They are:

\begin{enumerate}
	\item Computational theory: What is the goal of the computation, why is it appropriate, and what is the logic of the strategy by which it can be carried out?
	
	\item Representation and algorithm: How can this computational theory be implemented? In particular, what is the representation for the input and output, and what is the algorithm for the transformation?
		
	\item Hardware implementation: How can the representation and algorithm be realized physically? 
\end{enumerate}

When performing an information-theoretic analysis of a system one naturally faces the fundamental problem related to the coding of the information: In order to calculate (i.e.\ estimate) information theoretic quantities, one has to define a family of probability distributions over the state space of the system, each member of that family describing one system state that is to be considered. As we know, all information-theoretic quantities such as entropy and mutual information (between the system state and the state of some external quantity) are determined by the probability distributions involved. The big question now is how to define the system state in the first point, a question which is especially difficult to answer in the case of neural systems on all scales.

One possible way to construct such a probabilistic model for a sensory neurophysiological experiment involving just one neuron is the following. Typically, the experiment consists of many trials, where per trial $i=1,\dots,n$ in some defined time window a stimulus $S_i$ is presented eliciting a neural response $R(S_i)$ consisting of a sequence of action potentials. Presenting the same stimulus $S$ many times allows for the definition of a probability distribution of responses $R(S)$ of the neuron to a stimulus $S$. This is modeled as a conditional probability distribution $P_{R|S}$. As noted earlier, we usually have no direct access to $P_{R|S}$ but rather have to find an estimate $\hat{P}_{R|S}$ from the available data. Note that in practice, usually the joint probability distribution $P(R,S)$ is estimated and estimates of conditional probability distributions are subsequently obtained from the estimate of the joint distribution.

Let us now assume that the stimuli are drawn from the set of stimuli $S=\{S_1,\dots,S_k\}$ according to some probability distribution $P_S$ (that can be freely chosen by the experimenter). We can then compute the mutual information between the stimulus ensemble $S$ and its elicited response $R(S)$

\begin{equation*}
I(S;R(S)):=H(R(S)|S)-H(S)=H(S|R(S))-H(R(S))
\end{equation*}

using the probability distributions $P_S$ and $\hat{P}_{R|S}$, see Section~\ref{sec:mutualinformation}.

As usual, by mutual information we assess the expected shared information between the stimulus and its elicited response averaged over all stimuli and responses. In order to break this down to the level of single stimuli we can either consider the point-wise mutual information or employ one of the proposed decompositions of mutual information such as \emph{stimulus-specific information} or \emph{stimulus-specific surprise}, see \cite{Butts2003} for a review.


Having sketched the general setting let us come back to the question of coding of information by the neurons involved. This is important as we have to adjust our model of the neural responses accordingly, the goal being to capture all relevant features of the neural response in the model.  

Regarding neural coding, there are two main hypotheses of how single neurons might code information: Neurons could use a \emph{rate code}, i.e.\ encode the information via their mean firing rates neglecting the timing patterns of spikes or they could employ a \emph{temporal code}, i.e.\ a code where the precise timing of single spikes plays an important role. Yet another hypothesis would be that neurons code information in bursts of spikes, i.e.\ groups of spikes emitted in a small time window, which is a variant of the time code. For questions regarding coding in populations see the review \cite{Quiroga2009}.

Note that the question of neural coding is a highly debated one in the neurosciences as of today (see \cite{Sejnowski1995, Gerstner1997}) and we do not want to favor one view point over the other in the following. As with many things in nature there does not seem to be a clear black an white picture regarding neuronal coding. Rather it seems that a gradient of different coding schemes is employed depending on which sensory system is considered 
and at which stage of neuronal processing, see \cite{Gerstner1997, Buracas1998, Panzeri2001, Chechik2006, Rolls2011}.

\subsection{Computing Entropies of Spike Trains}

Let us now compute the entropy of spike trains and subsequently single spikes, assuming that the neurons we model either employ a rate or a time code. We are especially interested in the maximal entropy attainable by our model spike trains as these can give us upper bounds for the amount of information such trains and even single spikes can carry in theory. The following examples here are adapted from \cite{Trappenberg2010}. Concerning the topics of spike trains and their analysis, the interested reader is also pointed to \cite{Rieke1999}.

\begin{figure}
\centering
	\includegraphics{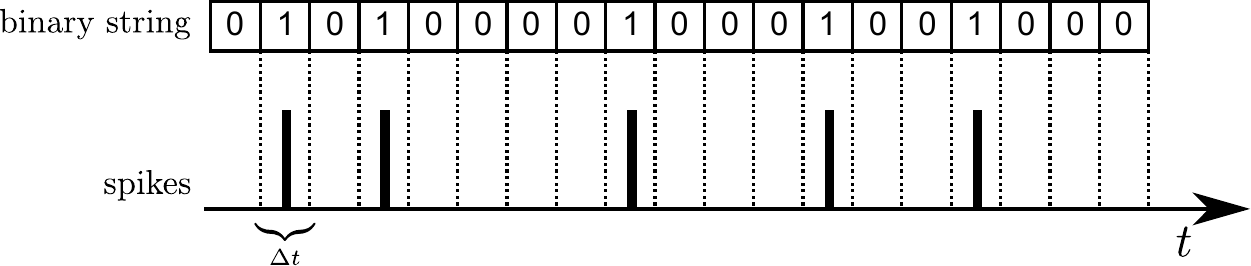}
\caption{Model of a spike train. The binary string is obtained through a binning of time.}
\label{fig:spikebins}
\end{figure}

First, we define a model for the spike train emitted by a neuron measured for some fixed time interval of length $T$. We can consider two different models for the spike train, a continuous and a discrete one. In the continuous case, we model each spike by a Dirac delta function and the whole spike train as a combination of such functions. The discrete model is obtained from the continuous one by introducing small time bins of size $\Delta t$ in a way that one bin can at most contain one spike, say $\Delta t = 2$ ms. We then assign to each bin in which no spike occurred a value of $0$ and ones in which a spike occurred a value of $1$, see Figure~\ref{fig:spikebins}.

Let us use this discrete model for the spike train of a neuron, representing a spike train as a binary string $S$ in the following. Fixing the time span to be $T$ and the bin width to be $\Delta t$, each spike train $S$ has length $N=T/\Delta t$. We want to calculate the maximal entropy among all such spike trains $S$, subject to the condition that the number of spikes in $S$ is a fixed number $r\leq N$ which we call the spike rate of $S$.   

Let us now calculate the entropy in the firing pattern of a neuron of which we assume that spike timing carries important information, i.e.\ a neuron employing a time code. In order to keep the model simple, let us further assume that the spiking behavior is not restricted in any way, i.e.\ that all possible binary strings $S$ are equiprobable. Then we can calculate the entropy of this uniform probability distribution $P$ as

\begin{equation}
H(P)=\log \binom{N}{r},
\label{eq:entropytimecode}
\end{equation}

where $\binom{N}{r}$ denotes the binomial coefficient $\binom{N}{r}=\frac{N!(N-r)!}{r!}$, the number of all distinct binary strings of length $N$ having exactly $r$ non-zero entries. The entropy in Equation~\ref{eq:entropytimecode} can be approximated by 

\begin{equation}
H(P)\approx -\frac{N}{\ln 2} \left( \frac{N}{r} \ln \frac{N}{r} +(1-\frac{N}{r})\ln ( 1 - \frac{N}{r}) \right),
\label{eq:entropytimeln}
\end{equation}

where $\ln$ denotes the natural logarithm to the base $e$. The expression in Equation~\ref{eq:entropytimeln} is obtained by using the approximation formula 

\begin{equation*}
\log \binom{n}{k} \sim n\left(\frac{k}{n}\log(\frac{k}{n}) -(1-\frac{k}{n})\log(1-\frac{k}{n})\right) 
\end{equation*}

which is valid for large $n$ and $k$ and in turn based on Stirling's approximation formula for $\ln n!$. 

\begin{figure}
\centering
	\includegraphics{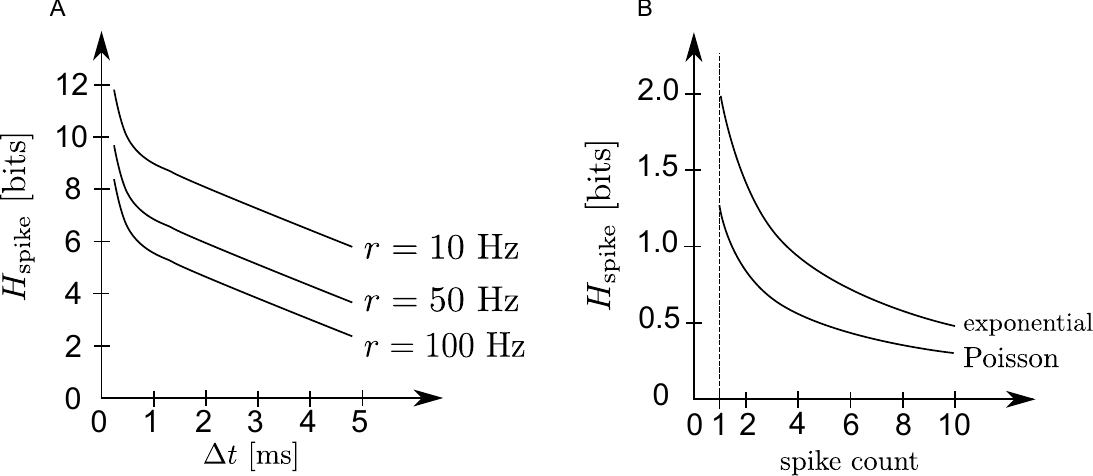}
\caption{Maximum entropy per spike for spike trains. (A) Time code with different rates $r$ as a function of the size $\Delta t$ of the time bins. (B) Rate code using Poisson and exponential spiking statistics. Figure adapted from \cite{Trappenberg2010} Fig. D.4.}
\label{fig:entropyrate}
\end{figure}

See Figure~\ref{fig:entropyrate}A for the maximum entropy attainable by the time code as a function of bin size $\Delta t$ for different firing rates $r$.  

On the other hand, modeling a neuron that reacts to different stimuli with a graded response it its firing rate is usually done using a rate code. Assuming a rate code where the timing of spikes does not play any role yields different results, as we will see in the following, see Figure~\ref{fig:entropyrate}B. In the rate code only the number of spikes $N$ occurring in a given time interval of length $T$ matters, i.e.\ we consider probability distributions $P_{N,T}$ parametrized by $N$ and $T$ describing how likely the occurrence of $N$ spikes in a time window of length $T$ is. Being well-backed with experimental data \cite{Softky1993, Compte2003, Nawrot2008}, a popular choice of $P_{N,T}$ is taking a Poisson distribution with some fixed mean $N=r\cdot T$, where $r$ is thought of as the mean firing rate of the neuron.  

The probability $P_{N,T}(n)$ of observing $n$ spikes in an interval of length $T$ now is given by the pmf of the Poisson distribution 

\begin{equation*}
P_{N,T}(n)=\frac{N^n e^{-N}}{n!}
\end{equation*}

and the entropy of $P_{N,T}$ computes as 

\begin{equation*}
H(P_{N,T})=-\sum_{n} P_{N,T}(n) \log P_{N,T}(n).
\end{equation*}

Again using Stirling's formula this can be written as

\begin{equation}
H(P_{N,T})\approx\frac{1}{2}\left(\log N - \log 2\pi \right).
\label{eq:entropypoissonrate}
\end{equation}

Dividing the entropy $H(P_{N,T})$ by the number of spikes that occurred yields the entropy per spike. See Figure~\ref{fig:entropyrate}B for a plot of the entropy per spike as a function of the number of observed spikes. 

An interesting question is to ask for the maximal information (i.e.\ entropy) that spike trains can carry, assuming a rate code. Assuming continuous time and prescribing mean and variance of the firing rate, this leaves the exponential distribution $P_\mathrm{exp}$ as the one with the maximal entropy. The entropy of an exponentially distributed spike train with mean rate $r=1/T(e^{\lambda}-1)$ is

\begin{equation*}
	H(P_\mathrm{exp})\approx\log (1+N) + N \log (1+ \frac{1}{N}),
\end{equation*}

see also Figure~\ref{fig:entropyrate}B.


Note that while it was possible to compute the exact entropies in the preceding as we assumed full knowledge of the underlying probability distributions. This is of course not the case for data obtained by recordings. Here the estimation of entropies faces the bias-related problems of sparsely sampled probability distributions as discussed earlier. Concerning entropy estimation in spike trains the reader is also pointed to \cite{Panzeri2007}.

\subsection{Efficient Coding?}

The principle of efficient coding \cite{Attneave1954, Barlow1961, Simoncelli2001} (also called \emph{Infomax principle}) was first proposed by Attneave and Barlow. It views the early sensory pathway as a channel in Shannon's sense and postulates that early sensory systems try to maximize information transmission under the constraint of an efficient code, i.e.\ that neurons maximize mutual information between a stimulus and their output spike train, using as few spikes as possible. This minimization of spikes for a given stimulus results in a maximal compression of the stimulus data, minimizing redundancies between different neurons on a population level. One key prediction of this optimality principle is that neurons involved in the processing of stimulus data (and ultimately the whole brain) is adapted to natural stimuli, i.e. some form of natural (and structured) sensory input such as sounds or images rather than noise. For some sensory systems it could be shown that there is strong evidence that early stages of processing indeed perform an optimal coding, see e.g.\ \cite{Olshausen1997}. While first mainly the visual system was studied and it was shown that the Infomax principle holds here \cite{Barlow1961}, other sensory modalities were also considered in the following years \cite{Bialek1989, Linsker1988, Linsker1990, Linsker1992, Hateren1992, Zhaoping2006}.

But whereas the Infomax principle could explain certain experimental findings in the early sensory processing stages, the picture becomes less clear the more upstream the information processing in neural networks is considered. Here, other principles were also argued for, see for example \cite{Globerson2009}.

On the system-level, Friston et al. \cite{Friston2006, Friston2010} proposed an information theoretic measure of free energy in the brain, that can be understood as generalization of the concept of efficient coding. Also arguing for optimal information transfer, Norwich \cite{Norwich1993} gave a theory of perception based on information-theoretic principles. He 
argues that the information present in some stimulus is relayed to the brain by the sensory system with negligible loss. Many empirical equations of psychophysics can be derived from this model. 

\subsection{Scales}

There are many scales at which information-theoretic analyses of neural systems can be performed. From the level of a single synapse \cite{Steveninck1996, London2002} over the level of single neurons \cite{Steveninck1988, Rolls2011} over the population level \cite{Pouget2000, Quiroga2009, Ince2010a, Crumiller2011, Fairhall2012} up to the system level \cite{Vicente2011, Ostwald2011}. In the former cases the analyses are usually carried out on electrophysiologically recorded data of single cells, whereas on the system level data is usually obtained by EEG, fMRI or MEG measurements.

Notice that most of the information-theoretic analyses of neural systems were done for early stages of sensory systems, focusing on the assessment of the amount of mutual information between some stimulus and its neural response. Here different questions can be answered, about the nature and efficiency of the neural code and the information conveyed by neural representations of stimuli, see \cite{Bialek1991, Rieke1993, Borst1999, Rolls2011}. This stimulus-response-based approach has already provided a lot of insight into the processing of information in early sensory systems, but things get more and more complicated the more downstream an analysis is performed \cite{Chechik2006, Rolls2011}.

On the systems level, the abilities of neural systems to process and store information are due to interactions of neurons, populations of neurons and sub-networks. As these interactions are highly-nonlinear and in contrast to the early sensory systems neural activity is mainly driven by the internal network dynamics (see \cite{Vicente2011, Arnold2013}), stimulus-response-type models often are not very useful here. Here, transfer entropy has proven to be a valuable tool here, making analyses of information transfer in the human brain in vivo possible \cite{Vicente2011, Ostwald2011}. Transfer entropy can also be used as a measure for causality, as we will discuss in the next section.











\subsection{Causality in the neurosciences}
\label{sec:causality}

The idea of \emph{causality}, namely the question of what are the causes resulting in the observable state and dynamics of complex systems of physical, biological or social nature is a deep, philosophical question that has been driving scientists in all fields ever since. In a sense this question lies at the heart of science itself and as such is often notoriously difficult to answer.

In the neurosciences, this principle is related to one of the core questions of neural coding and subsequently neural information processing: What stimuli make neurons spike (or change their membrane potential, for non-spiking neurons)? For many years now, neuroscientists have investigated  neurophysiological correlates of information presented to a sensory system in form of stimuli. 

While considerable progress has been made regarding the answer to this question in the early stages of sensory processing (see the preceding sections), where often a clear correlation between a stimulus and the resulting neuronal activity could be found, things get less and less clear the further downstream this question is addressed. In the latter case, neuronal activity is subject to higher and higher degrees of internal dynamics and a clear stimulus-response relation is often lacking. 

Considering early sensory systems, even though merely a correlation between a stimulus and neural activity can be measured, it is justified to speak of causality here, as it is possible to actively influence the stimulus and observe the change in neural activity. Note that the idea of intervention is crucial here, see \cite{Pearl2000, Ay2008}.

Looking at more downstream systems or at the cognitive level, an active intervention albeit possible (but often not as directly as for sensory systems) may not have the same easy to detect effects on system dynamics. Here, often just statistical correlations can be observed and in most cases it is very hard if not impossible to show that the principle of causality in its purest form holds. Yet, one can still make some statements regarding what one might call ``statistical causality'' in this case, as we will see. 


In an attempt to give a statistical characterization of the notion of causality, the mathematician Wiener \cite{Wiener1956} came up with the following probabilistic framing of this concept that came to be known as \emph{Wiener causality}: Consider two stochastic processes $X=(X_t)_{t\in \mathbb{N}}$ and $Y=(Y_t)_{t\in \mathbb{N}}$. Then $Y$ is said to Wiener-cause $X$ if the knowledge of past values of $Y$ diminishes uncertainty about the future values of $X$. Note that Wiener causality is a measure of predictive information transfer and not one of causality and thus the naming is a bit unfortunate, see \cite{Lizier2010a}.


The economist Granger employed Wiener's principle of causality and developed the notion of what is nowadays called \emph{Wiener-Granger causality} \cite{Granger1969, Bressler2011}. Subsequently, the linear Wiener-Granger causality and its generalizations were often employed as measure of statistical causality in the neurosciences, see \cite{Hlavackovaschindler2007, Bressler2011}. Another model for causality in the neurosciences is \emph{dynamic causal modeling} \cite{Friston2003, Stephan2007, Friston2011}. 

In contrast to dynamic causal modeling, causality measures based on information-theoretic concepts are usually purely data-driven and thus inherently model-free \cite{Hlavackovaschindler2007, Vicente2011}. This fact can be of advantage in some cases but we do not want to make a judgment here, calling one method better per se, as each has its advantages and drawbacks \cite{Friston2012}. 

The directional and time-dynamic nature of transfer entropy allows using it as a measure of Wiener-causality, as was proposed in the field of neurosciences recently \cite{Vicente2011}. As such, transfer entropy can be seen as a non-linear extension of the concept of Wiener-Granger causality, see \cite{Lungarella2007} for an comparison of transfer entropy to other measures. 


Note again that transfer entropy still essentially is a measure of conditional correlation rather than one of direct effect (i.e.\ causality) and that correlation is not causation. Thus it is a philosophical question to which extent transfer entropy can be used to infer some form of causality, a question that we will not further pursue here, rather pointing the reader to \cite{Pearl2000, Hlavackovaschindler2007, Lungarella2007, Ay2008}. 


In any case the statistical significance of the inferred causality (remember that transfer entropy just measures conditional correlation) has to be verified. For trial based data-sets as often found in the neurosciences, this testing is usually done against the null-hypothesis $H_0$ of average transfer entropy obtained by random shuffling of the data.

\subsection{Information-theoretic Aspects of Neural Dysfunction}

Given the fact that information-theoretic analyses can provide insights about the functioning of neural systems, the next logical step is to ask how these this might help in better understanding neural dysfunction and neural diseases, maybe even giving hints to new treatments. 

The field one might call ``computational neuroscience of disease'' is an emerging field of research within the neurosciences, see the special issue of \emph{Neural Networks} \cite{Cutsuridis2011}. The discipline faces some hard questions as in many cases dysfunction is observed on the cognitive (i.e.\ systems-) level but has causes on many scales of neural function (sub-cellular, cellular, population, system).

Over the last years, different theoretical models regarding neural dysfunction and disease were proposed, among them computational models applicable to the field of psychiatry \cite{Huys2011, Montague2012}, models for brain lesions \cite{Alstott2009}, models of epilepsy \cite{IvanSoltesz2011}, models for deep brain stimulation \cite{McIntyre2007, Pirini2009}, models for aspects of Parkinson's \cite{Haeri2005, Moustafa2011}  and Alzheimer's \cite{Bhattacharya2011, Krishnamurti2011} disease, of abnormal auditory processing \cite{Du2011, Krishnamurti2011} and for congenital prosopagnosia (a deficit in face identification) \cite{Stollhoff2011}.


Some of these models employ information-theoretic ideas in order to assess differences between the healthy and dysfunctional states \cite{Stollhoff2011, Barcelo2007a}. For example, information-theoretic analyses of cognitive and systems-level processes in the prefrontal cortex were carried out recently \cite{Koechlin2007, Barcelo2007a} and differences in information-processing could be assessed between the healthy and dysfunctional system by means of information-theory \cite{Barcelo2007a}. 

Yet, computational neuroscience of disease is a very young field of research and it remains to be elucidated if and in what way analyses of neural systems employing information-theoretic principles could be of help in medicine on a broader scale.












\section{Software}
\label{sec:software}

There exist several open source software packages that can be used to estimate information-theoretic quantities of neural data. The list below is by no means complete, but should give a good overview of things, see also \cite{Ince2010b}.

\begin{itemize}

\item \texttt{entropy: Entropy and Mutual Information Estimation}\\
URL: \url{http://cran.r-project.org/web/packages/entropy}\\
Authors: Jean Hausser and Korbinian Strimmer.\\
Type: R package\\
From the website: This package implements various estimators of entropy, such as the shrinkage estimator by Hausser and Strimmer, the maximum likelihood and the Millow-Madow estimator, various Bayesian estimators, and the Chao-Shen estimator. It also offers an R interface to the NSB estimator. Furthermore, it provides functions for estimating mutual information.

\item \texttt{information-dynamics-toolkit}\\
URL: \url{http://code.google.com/p/information-dynamics-toolkit}\\
Author: Joseph Lizier\\
Type: standalone Java software\\
From the website: Provides a Java implementation of information-theoretic measures of distributed computation in complex systems: i.e. information storage, transfer and modification. Includes implementations for both discrete and continuous-valued variables for
entropy, entropy rate, mutual information, conditional mutual information,
transfer entropy, conditional/complete transfer entropy, active information storage, excess entropy / predictive information, separable information. 	

\item \texttt{ITE (Information Theoretical Estimators)}\\
URL: \url{https://bitbucket.org/szzoli/ite/}\\
Author: Zoltán Szabó\\
Type: Matlab/Octave plugin\\
From the website: ITE is capable of estimating many different variants of entropy, mutual information and divergence measures. Thanks to its highly modular design, ITE supports additionally the combinations of the estimation techniques, the easy construction and embedding of novel information theoretical estimators, and  their immediate application in information theoretical optimization problems. ITE can estimate Shannon-, Rényi entropy; generalized variance, kernel canonical correlation analysis, kernel generalized variance, Hilbert-Schmidt independence criterion, Shannon-, L2-, Rényi-, Tsallis mutual information, copula-based kernel dependency, multivariate version of Hoeffding's Phi; complex variants of entropy and mutual information; L2-, Rényi-, Tsallis divergence, maximum mean discrepancy, and J-distance. ITE offers solution methods for Independent Subspace Analysis (ISA) and its extensions to different linear-, controlled-, post nonlinear-, complex valued-, partially observed systems, as well as to systems with nonparametric source dynamics.

\item \texttt{PyEntropy}\\
URL: \url{http://code.google.com/p/pyentropy}\\
Authors: Robin Ince,  Rasmus Petersen, Daniel Swan, Stefano Panzeri\\
Type: Python module\\
From the website: pyEntropy is a Python module for estimating entropy and information theoretic quantities using a range of bias correction methods.

\item \texttt{Spike Train Analysis Toolkit}\\
URL: \url{http://neuroanalysis.org/toolkit}\\
Authors:  Michael Repucci, David Goldberg, Jonathan Victor, Daniel Gardner\\
Type: Matlab/Octave plugin\\
From the website: Information theoretic methods are now widely used for the analysis of spike train data. However, developing robust implementations of these methods can be tedious and time-consuming. In order to facilitate further adoption of these methods, we have developed the Spike Train Analysis Toolkit, a software package which implements several information-theoretic spike train analysis techniques.

\item \texttt{TRENTOOL}\\
URL: \url{http://trentool.de}\\
Authors: Michael Lindner, Raul Vicente, Michael Wibral, Nicu Pampu and Patricia Wollstadt\\
Type: Matlab plugin\\
From the website: TRENTOOL uses the data format of the open source MATLAB toolbox Fieldtrip, that is popular for electrophysiology data (EEG/MEG/LFP).
Parameters for delay embedding are automatically obtained from the data. TE values are estimated by the Kraskov-Stögbauer-Grassberger estimator and subjected to a statistical test against suitable surrogate data. Experimental effects can then be tested on a second level. Results can be plotted using Fieldtrip layout formats.

\end{itemize}

\section{Acknowledgements}

The author would like to thank \emph{Nihat Ay}, \emph{Yuri Campbell}, \emph{Jörg Lehnert}, \emph{Timm Lochmann}, \emph{Wiktor Młynarski} and \emph{Carolin Stier} for their useful comments on the manuscript.


\end{document}